\documentclass[conference]{IEEEtran}
\IEEEoverridecommandlockouts

\usepackage[hyphens]{url}
\usepackage[pdftex,hidelinks]{hyperref}
\usepackage{amsmath,amsthm}
\usepackage{algorithm}
\usepackage{algpseudocodex}
\usepackage{multirow}
\usepackage{enumitem}
\usepackage{subcaption}
\usepackage{wasysym}
\usepackage{xspace}
\usepackage[table]{xcolor}
\usepackage{booktabs}
\usepackage[hyphens]{url}
\usepackage{amsthm}
\usepackage{amsmath,amsfonts,bm}
\usepackage{tcolorbox}
\usepackage{makecell}
\usepackage{url}

\newcommand{\lossfunc}[1]{\nabla_{#1}\loss}

\newcommand{\gradRef}[1]{\gradient{#1}{ref}}
\newcommand{\gradFin}{\gradient{}{final}}

\newcommand{\fedwm}{\textsc{CamoMark}\xspace}

\newcommand{\embedsCurr}{\mathbf{E}_b}
\newcommand{\embedsOrig}{\mathbf{E}_b^0}
\newcommand{\embedsBest}{\mathbf{E}_b^\star}

\newcommand{\loss}{\mathcal{L}}
\newcommand{\lossRec}{\loss_{\text{rec}}}
\newcommand{\lossReg}{\loss_{\text{reg}}}

\newcommand{\lr}{\eta}
\newcommand{\lrServer}{\lr_{\text{S}}}
\newcommand{\lrClient}{\lr_{\text{C}}}
\newcommand{\lrEmbedding}{\lr_{\text{E}}}
\newcommand{\coeReg}{\lambda}

\newcommand{\optServer}{\textsc{ServerOpt}\xspace}
\newcommand{\optClient}{\textsc{ClientOpt}\xspace}

\newcommand{\costG}{C_g}
\newcommand{\costH}{C_h}

\theoremstyle{definition}                               

\newcommand{\myparatight}[1]{\noindent{\bf {#1}.}~}

\newcommand{\numclients}{N}
\newcommand{\setclients}{C}

\newcommand{\globalM}[1]{\mathcal{M}_\params^{#1}}
\newcommand{\genM}{\mathcal{M}_{\omega}}

\newcommand{\params}{\theta}
\newcommand{\globalparams}[1]{\params^{#1}}
\newcommand{\localparams}[2]{\params_{#1}^{#2}}
\newcommand{\referenceparams}[2]{%
  \ifx\relax#1\relax
    \params_{#2}^{ref}%
  \else
    \params_{#1}^{#2,ref}%
  \fi
}

\newcommand{\updates}[2]{\Delta\params_{#1}^{#2}}
\newcommand{\gradient}[2]{{g}_{#1}^{#2}}

\newcommand{\cleanupdates}{{C}_\Delta}
\newcommand{\allupdates}{{U}_\Delta}

\newcommand{\agg}{\textsc{Agg}}

\newcommand{\wmratio}{\epsilon}   
\newcommand{\wmworker}{k}

\newcommand{\cleanmean}{\mu_C}
\newcommand{\cleanvar}{\|\Sigma_C\|_2}
\newcommand{\coeff}{\beta}

\newcommand{\numlayer}{L}
\newcommand{\eachlayer}{\ell}
\newcommand{\wmworkerl}{\mathcal{W}_{\eachlayer}} 
\newcommand{\filterworkerl}{\mathcal{F}_{\eachlayer}} 
\newcommand{\wmfilterworkerl}{\wmworkerl \cap \filterworkerl}
\newcommand{\avgfilter}{\frac{1}{\numlayer} \sum_{\eachlayer=1}^{\numlayer}}
\newcommand{\evasionrate}{\mathrm{ER}}

\newcommand{\data}{D}
\newcommand{\wmdata}{D^{w}}

\newcommand{\batch}{b}
\newcommand{\batches}{\mathcal{B}}

\newcommand{\vocab}{\mathcal{V}}
\newcommand{\prompt}{\pi}
\newcommand{\outputs}{\textbf{x}}

\newcommand{\watermark}[2]{\textsc{Watermark}^{#1}_{#2}}
\newcommand{\detect}[2]{\textsc{Detect}^{#1}_{#2}}

\newcommand{\pval}{\textup{$p$-value}\xspace}
\newcommand{\pvals}{\textup{$p$-values}\xspace}

\newcommand{\secretkey}{s}
\newcommand{\logitsbias}{\delta}
\newcommand{\greenratio}{\gamma}

\newcommand{\ngrams}{k}

\newcommand{\nullhypo}{H_{0}}

\newcommand{\alpaca}{\textup{Alpaca}\xspace}
\newcommand{\openweb}{\textup{OpenWebText}\xspace}
\newcommand{\oasst}{\textnormal{OASST1}\xspace}

\newcommand{\Kirchenbauer}{\textup{KGW+}}
\newcommand{\Kuditipudi}{\textup{KTH+}}
\newcommand{\ITS}{\textup{KTH-ITS}}
\newcommand{\EXP}{\textup{KTH-EXP}}
\newcommand{\Aarson}{\textup{AAR}}

\newcommand{\pythia}{\textup{Pythia}\xspace}
\newcommand{\qwen}{\textup{Qwen}\xspace}
\newcommand{\llama}{\textup{LLaMA}\xspace}

\newcommand{\fedavg}{\textup{FedAvg}\xspace}
\newcommand{\fedopt}{\textnormal{FedOpt}\xspace}

\newcommand{\randeigen}{\textup{RandEigen}\xspace}
\newcommand{\filtering}{\textup{Filtering}}
\newcommand{\noregret}{\textup{No-Regret}}

\newcommand{\hidra}{\textnormal{HiDRA}\xspace}

\newtcolorbox{embox}{
    colback=white,
    coltext=black,
    boxrule=1pt,
    width=\linewidth,
}

\begin{document}
 
\title{Watermarking for Byzantine Robust LLM Training}

\author{
  \IEEEauthorblockN{Leixu Huang}
  \IEEEauthorblockA{
    Georgia Institute of Technology\\
    \texttt{lhuang340@gatech.edu}
  }
  \and
  \IEEEauthorblockN{Zedian Shao}
  \IEEEauthorblockA{
    Georgia Institute of Technology\\
    \texttt{zedian.shao@gatech.edu}
  }
  \and
  \IEEEauthorblockN{Teodora Baluta}
  \IEEEauthorblockA{
    Georgia Institute of Technology\\
    \texttt{teobaluta@gatech.edu}
  }
}

\maketitle

\begin{abstract}
\boldmath
LLM watermarking has emerged as a promising approach for detecting downstream model misuse in centralized settings, but extending it to federated learning (FL) presents unique challenges. In particular, FL systems often rely on strong robust aggregators to defend against poisoning attacks. We empirically show that a malicious server can exploit these same aggregators as effective watermark-removal mechanisms during training, causing representative existing LLM watermarking methods to fail detection in the final model.
To address this limitation, we formalize the problem as a constrained optimization task and propose \fedwm, an aggregator-agnostic watermarking scheme for embedding robust watermarks in FL. \fedwm optimizes watermarked updates to become inliers, maintaining bounded deviation from the distribution of clean updates, while preserving sufficient watermark signal for reliable detection. Evaluated on \qwen fine-tuned on the \alpaca dataset, \fedwm achieves a detection confidence of $\pval = 1.82\times10^{-2}$ using only 3.3\% watermarked data. The watermark also remains robust to a range of post-training removal attacks, including continued fine-tuning on clean data, pruning, and model distillation. At higher watermark ratios, detection confidence improves substantially, reaching $\pval = 5.11\times10^{-11}$ at a 13.3\% watermark ratio.

\unboldmath
\end{abstract}

\section{Introduction}

Large language models (LLMs) are typically fine-tuned for alignment and performance on specialized downstream tasks~\cite{ouyang2022training,wei2021finetuned}. The fine-tuning datasets driving these advances are increasingly valuable but often distributed across multiple sources, such as different organizations, institutions, or private datasets~\cite{li2021survey,kairouz2021advances}. Federated learning (FL) has emerged as a decentralized paradigm for collaboratively training LLMs across multiple data owners~\cite{wu2025survey,rieke2020future,zhang2024towards,kairouz2021advances}. In FL, data owners (called clients) train the LLM locally on their data and only share model updates with a central server~\cite{mcmahan2017communication, kairouz2021advances, rieke2020future}. The server aggregates these updates into a global model, which is then sent back to the clients for further training. 

FL captures a setting in which clients explicitly consent to contribute their data for training the model. However, existing FL frameworks provide no guarantees about how the resulting global model will be used {\em after training}, leaving clients vulnerable to post-training misuse or unauthorized re-purposing. Growing evidence suggests that fine-tuned models may be repurposed for uses that contradict training data's original intent such as being adapted for military application~\cite{assaad2024meta} or used to create competing derivative works that undermine the original data's economic value~\cite{longpre2023dataprovenanceinitiativelarge,longpre2024data}.

When model misuse happens, clients require a mechanism to detect whether a suspect model has been trained on their data given only black-box access to the model. A common approach in prior work is to inject backdoors or trigger-based watermarks in the training data~\cite{adi2018turning,yang2024fedgmark,li2022fedipr,shao2024fedtracker,jia2021entangled,li2023black}. However, these require mislabeled data~\cite{yang2024fedgmark}, a shared secret across participants for verification~\cite{yang2024fedgmark}, and are limited to classification models~\cite{li2022fedipr,yang2024fedgmark,shao2024fedtracker}.

Recent work in the centralized setting uses watermarks embedded in LLM-generated data to detect whether other language models were fine-tuned on it~\cite{sander2024watermarking}.
LLM watermarking plants statistically detectable signals into model-generated text during the decoding process, while keeping the output distribution as close as possible to that of non-watermarked generations~\cite{aaronson2023watermarking,kirchenbauer2023watermark,kuditipudi2023robust}. 
This enables high-confidence detection after training on LLM-generated watermarked data, i.e., $\pval \leq 10^{-75}$~\cite{gu2023learnability}. Such small \pval indicates that the observed watermark signal is unlikely to arise if the examined LLM never trained on watermarked data~\cite{cui2025robust,wei2024proving,sander2024watermarking,gu2023learnability}. Even after further fine-tuning on non-watermarked data, detection remains highly reliable~\cite{sander2024watermarking}. To the best of our knowledge, though, no prior work has studied whether LLM watermarks remain detectable after FL.

Compared to the centralized setting, federated setups present a unique challenge that the server can actively filter out anomalous updates. Since most existing watermarks rely on mislabeled data~\cite{adi2018turning,yang2024fedgmark} or fictitious knowledge~\cite{cui2025robust}, they closely resemble poisoned updates from Byzantine clients. A principled approach to remove such updates is via \textbf{robust mean estimation}~\cite{diakonikolas2019recent, blanchard2017machine, yin2018byzantine}. In particular, strong robust aggregators iteratively identify and remove the most deviant client updates until the distance between the true mean of clean updates and the aggregated output is optimally bounded~\cite{diakonikolas2017being, hopkins2020robust, kothari2017outlier, choudhary2024attacking,lee2025practical}. While LLM watermarks are designed to embed statistical signals close to the original LLM generations, we show that the server can still deploy robust aggregator as a removal attack to filter these watermarked updates. This naturally raises the question: \emph{Can we design a watermarking approach that remains detectable via black-box access, while evading strong robust aggregators?}

The core challenge in designing a robust watermark for FL is that watermark detectability is often in contradiction to surviving the strong robust aggregator. High detectability makes watermarked updates deviate significantly from the clean mean, precisely what the robust aggregators are designed to eliminate. To resolve this tension, we observe that detectability does not require a large deviation; rather, it requires a consistent one. By constraining watermarked updates to remain within a bounded deviation from the clean mean, they survive robust aggregation. Although the watermark signal is small at each step, as the signal is accumulated across rounds, it remains statistically detectable downstream.

We propose \fedwm, a watermarking algorithm that crafts watermarked updates to evade removal by strong robust aggregators while preserving detectability. We formulate the objective of \fedwm as a constrained optimization problem in the embedding space. Specifically, we keep the watermarked text fixed, because the watermark signal is embedded at the token-level. Directly optimizing over discrete input tokens is not only computationally difficult, but may yield different texts across rounds. Changing tokens at any point during training is equivalent to a self-inflicted paraphrasing attack, making post-training detection less effective. We instead optimize the continuous input embeddings of the watermarked samples to minimize the distance between their gradients and clean proxy gradients, increasing the probability of watermarked updates evading robust aggregators. Moreover, we find that the resulting optimized embeddings retain substantial semantic information and reliably map back to the original token sequence. Therefore, the watermark signal is preserved, enabling high-confidence detection post training. 

We evaluate $\fedwm$ across three models (\pythia~\cite{biderman2023pythia}, \qwen~\cite{bai2023qwen}, and \llama~\cite{touvron2023llama}) and two datasets (\alpaca~\cite{taori2023stanford} and \openweb~\cite{Gokaslan2019OpenWeb}) used in prior centralized work~\cite{gu2023learnability, sander2024watermarking}. In all settings, $\fedwm$ achieves high-confidence watermark detection even when the server actively attempts to remove the watermark using the state-of-the-art aggregator $\randeigen$~\cite{lee2025practical} during training.  Evaluated on \qwen fine-tuned on the \alpaca dataset, \fedwm achieves a detection confidence of $\pval = 1.82\times10^{-2}$ using only 3.3\% watermarked data. At higher watermark ratios, detection confidence improves substantially, reaching $\pval = 5.11\times10^{-11}$ at a 13.3\% watermark ratio. We further evaluate post-training watermark removal attacks on the fine-tuned global model, including additional fine-tuning on clean datasets~\cite{sander2024watermarking, hu2024stable}, pruning~\cite{liu2018rethinking, renda2020comparing, chen2022copy}, and distillation~\cite{yang2024fedgmark, shafieinejad2021robustness}. These scenarios capture attempts by any malicious FL participant, e.g., the server or a malicious client, trying to remove the watermark before release. We evaluate these attacks under the constraint that the adversary preserving model utility, and show that \fedwm remains reliably detectable across most evaluated settings.

\myparatight{Summary of Contributions} To the best of our knowledge, this is the first work to introduce federated LLM fine-tuning data detection. In summary, our contributions are:
\begin{itemize}
    \item \textbf{New Threat Model.} We show that watermark detectability observed in centralized fine-tuning extends to FL when the server aggregates client updates by averaging. We formalize an in-training adversary threat model for FL model ownership and demonstrate that a malicious server can effectively remove watermark signals using the state-of-the-art practical, strong Byzantine-robust aggregator, $\randeigen$. Other strong robust aggregators that cannot achieve their theoretical bias bound in practice are not effective adversaries in our threat model.

    \item \textbf{Algorithm.} We propose an aggregator-agnostic watermarking algorithm $\fedwm$ that is robust in FL setup. We formulate it as a constrained optimization that shapes watermarked updates to be within the statistically optimal bias bound per-round and align directionally with clean updates. The design thus is not contingent on a particular aggregator and helps evade practical Byzantine-robust aggregators, while preserving detectability.
 
    \item \textbf{Experimental.} We evaluate our approach across two datasets and three models from distinct families. $\fedwm$ enables high-confidence black-box detection ($p$-value $<$ $10^{-10}$) even in the presence of strong robust aggregators and remains resilient to post–training watermark-removal attacks.
    
\end{itemize}

\section{Motivation and Background}
\begin{figure*}[t]
    \centering
    \begin{minipage}{0.51\linewidth}
        \centering
            \includegraphics[width=\linewidth]{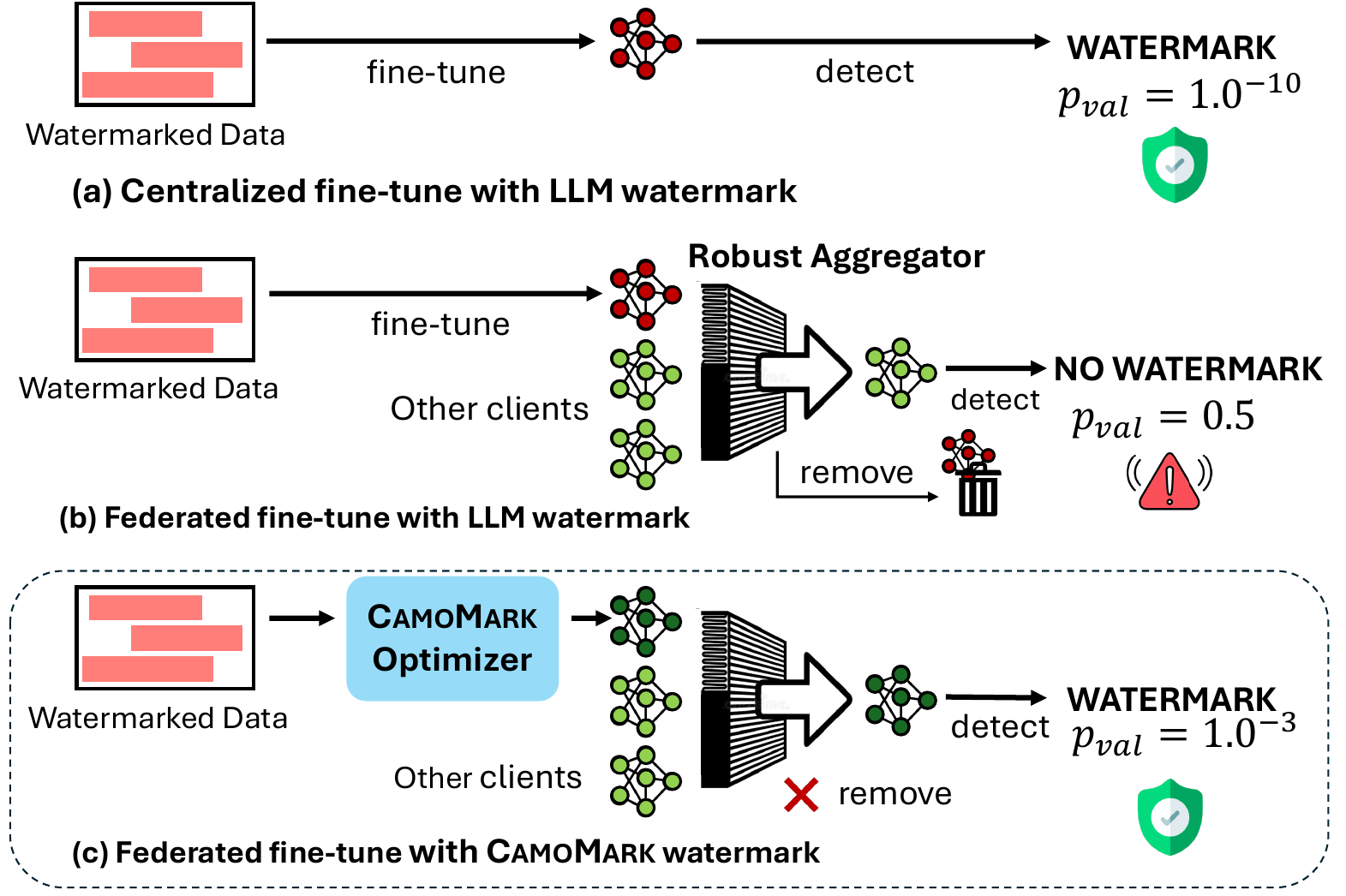}
        \caption{LLM watermark detectability under centralized and federated fine-tuning. In FL, a malicious server can use strong robust aggregation as an in-training watermark removal attack.}
        \label{fig:threat}
    \end{minipage}
    \hfill
    \begin{minipage}{0.44\linewidth}
        \centering
        \includegraphics[width=\linewidth]{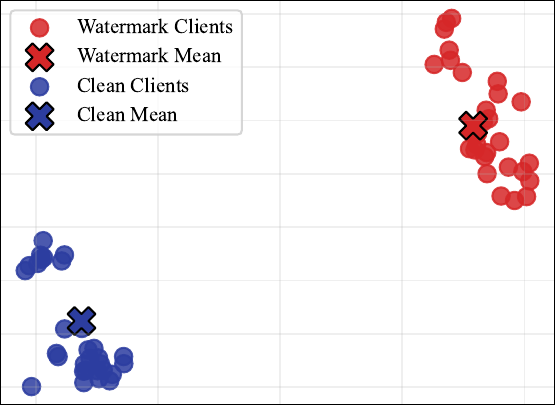}
        \caption{t-SNE visualization of $\Kirchenbauer$ updates (red) and clean synthetic updates (blue). $\Kirchenbauer$ updates form visually distinct clusters from clean updates, even when both datasets are generated by the same model}
        \label{fig:updates_diff}
    \end{minipage}
\end{figure*}

\subsection{Motivation}
\label{sec:motivation}
Model developers increasingly emphasize that training data should be used only with proper consent and under agreed deployment conditions~\cite{longpre2023dataprovenanceinitiativelarge, gebru2021datasheets, longpre2024consent}. In practice, however, those conditions are not always upheld after training. Data owners may permit model training for one purpose, only to later discover that the resulting model was deployed for another. For example, Llama 2's Acceptable Use Policy explicitly lists military use as prohibited, yet Meta later granted the U.S. government and its contractors access for national security purposes~\cite{assaad2024meta}. MPT-StoryTeller~\cite{frankle2023tweet} was fine-tuned on Books3~\cite{gao2020pile} and initially released under a commercial license before undergoing license revisions following copyright concerns~\cite{longpre2023dataprovenanceinitiativelarge}. WizardCoder~\cite{luo2023wizardcoder} used a restricted OpenAI dataset but was released under a commercial license~\cite{longpre2023dataprovenanceinitiativelarge}.

FL makes this misuse harder to attribute and easier to hide. Unlike centralized training, FL distributes training across many clients that may not share governance or visibility into downstream deployment~\cite{kairouz2021advances, lyu2020towards}. Prior work has mostly studied misuse from the server's perspective, where the concern is malicious clients~\cite{bhagoji2019analyzing, 247652, yang2024fedgmark}. Here we focus on the reverse direction: the server itself may later deploy the final global model in an unauthorized downstream setting. Benign clients therefore need an independent way to test whether a suspect model was fine-tuned on their data.

In this paper, we ask whether LLM-generated watermarked data can support reliable post-training detection in FL when the server uses Byzantine-robust aggregation during training.

\subsection{Background}

We briefly review the FL and watermarking concepts used in our setup.

\subsubsection{Federated Setup}
We consider federated fine-tuning with $\numclients$ clients $\setclients = \{c_1, \dots, c_N\}$ and global model $\globalM{}$ with parameters $\params \in \mathbb{R}^d$. In round $t$, the server broadcasts $\globalparams{t}$ to all clients. Each client $c_i$ fine-tunes on its private dataset $\data_i$ and returns a local update $\updates{i}{t}$. The server aggregates the updates via $\agg: \mathbb{R}^{\numclients \times d} \rightarrow \mathbb{R}^d$:
\[
    \globalparams{t+1} = \globalparams{t} + \agg\left(\{\updates{i}{t}\}_{i=1}^N\right).
\]

\subsubsection{Byzantine Robust Aggregators}
Byzantine-robust aggregators estimate the mean of clean client updates through limiting the influence of corrupted ones~\cite{diakonikolas2019recent, blanchard2017machine, yin2018byzantine}. Let $\allupdates$ denote the full set of $\numclients$ clients updates, of which a fraction $\wmratio$ may be arbitrarily corrupted. Let $\cleanupdates \subseteq \allupdates$ be the clean subset with mean $\cleanmean = \frac{1}{|\cleanupdates|} \sum_{\updates{i}{t} \in \cleanupdates} \updates{i}{t}$. A strong robust aggregator~\cite{diakonikolas2017being, hopkins2020robust, kothari2017outlier, choudhary2024attacking,lee2025practical} guarantees
\begin{align*}
    \left\| \agg(\allupdates) - \cleanmean \right\|_2 \leq \coeff \cdot \cleanvar^{\frac{1}{2}}
\end{align*}
where $\cleanvar$ is the maximum variance of $\cleanupdates$. We refer to the left-hand side as \textit{bias}: the gap between the aggregated output and $\cleanmean$. 

Strong robust aggregators achieve $\coeff = O(\sqrt{\wmratio})$ by iteratively filtering outliers until the remaining set satisfies the desired bias bound. We focus on strong robust aggregators because they achieve the statistically optimal bias guarantee, so any watermark that survives them will, by construction, also survive weaker aggregation rules.

\myparatight{Quasi-linear-time Strong Robust Aggregators}
For LLMs, computing the maximum variance across the full parameter dimension is prohibitively expensive. Strong robust aggregators~\cite{diakonikolas2017being, hopkins2020robust, kothari2017outlier} that require such computation cannot attain their theoretical bias guarantees at LLM scale. Practical strong robust aggregation is nonetheless an active and advancing direction. \randeigen~\cite{lee2025practical}, to our knowledge, is the state-of-the-art method that attains its theoretical bias guarantee in practice, representing the strongest practical instance of this class available to evaluate against today. We therefore focus on \randeigen in this paper. \randeigen projects updates onto random subspaces and filters based on those lower-dimensional projections. Its stopping criterion is based on convergence of variance estimates across iterations, not a fixed threshold. As a result, \randeigen always performs at least one filtering step even when the bias bound is already satisfied.

\subsubsection{LLM Watermarking}
Let $\genM$ map an input prompt $\prompt=(x_1,\ldots,x_q)\in\vocab^q$ of length $q$ over vocabulary $\vocab$ to an output sequence $\outputs\in\vocab^Q$ of length $Q$ through autoregressive decoding.

To watermark generations of $\genM$, $\watermark{\genM}{\secretkey}(\prompt) \rightarrow \outputs^w$ uses a secret key $\secretkey$ to perturb decoding and produce a watermarked response $\outputs^w$~\cite{zhao2025sok}. A detector $\detect{}{\secretkey}(\outputs) \rightarrow \pval$ then tests whether a sequence contains the watermark signal. Smaller \pval indicates stronger evidence that the sequence is watermarked. Existing methods are often divided into low-distortion~\cite{kirchenbauer2023watermark} and distortion-free approaches~\cite{aaronson2023watermarking,kuditipudi2023robust}. The former explicitly biases decoding, while the latter  preserves the original output distribution.

\myparatight{Detectability after fine-tuning} Let $\genM$ generate a watermarked dataset $\wmdata$. To test whether another model $\globalM{}$ was fine-tuned on $\wmdata$, we use $\detect{\globalM{}}{\secretkey}(\wmdata) \rightarrow \pval$, which queries $\globalM{}$ on $\wmdata$, computes a watermark score from its predictions~\cite{sander2024watermarking,gu2023learnability}, and evaluates the score with the hypothesis test:
\begin{align}
H_0 &: \globalM{} \text{ was not trained on } \wmdata \\
H_1 &: \globalM{} \text{ was trained on } \wmdata
\end{align}
Lower $\pval$ gives stronger evidence that $\globalM{}$ was trained on $\wmdata$. Throughout this paper, detectability refers to this post-fine-tuning notion unless otherwise specified.

\section{Problem Formulation}
\begin{table*}[t]
\centering
\caption{Watermark detectability under different federated learning aggregation schemes. \fedwm is the only watermark that remains fully detectable across all evaluated settings. Most in-process watermarks survive \fedavg~\cite{mcmahan2017communication}, \filtering~\cite{diakonikolas2017being}, and \noregret~\cite{hopkins2020robust}, but lose detectability under \randeigen~\cite{lee2025practical}, the state-of-the-art strong robust aggregator. $\CIRCLE$ = detectable; $\LEFTcircle$ = partially detectable; $\Circle$ = not detectable.}
\label{tab:summary}
\begin{tabular}{lll|c|ccc|cc}
\toprule
& & 
& \textbf{Post-hoc}
& \multicolumn{3}{c|}{\textbf{In-Process Watermark}}
& \multicolumn{2}{c}{\textbf{FL-Aware Watermark}} \\
& & 
& \textbf{Watermark}
& & & & \multicolumn{2}{c}{\textbf{(\textit{Ours})}} \\
\cmidrule(lr){4-4} \cmidrule(lr){5-7} \cmidrule(lr){8-9}
\multicolumn{2}{l}{\textbf{Aggregator}} & \textbf{Clean Data}
  & \textbf{ASCII}
  & \textbf{KTH-ITS} & \textbf{KGW+, AAR} & \textbf{KTH-EXP}
  & \textbf{Convex Combo.} & \textbf{$\fedwm$} \\
\midrule
\multicolumn{2}{l}{FedAvg} & —
  & $\Circle$ & $\Circle$ & $\CIRCLE$ & $\CIRCLE$ & $\CIRCLE$ & $\CIRCLE$ \\
\midrule
\multirow{4}{*}{\shortstack[l]{Strong Robust\\Agg.}}
  & \filtering & \multirow{2}{*}{—}
  & \multirow{2}{*}{$\Circle$} & \multirow{2}{*}{$\Circle$} & \multirow{2}{*}{$\CIRCLE$} & \multirow{2}{*}{$\CIRCLE$} & \multirow{2}{*}{$\CIRCLE$} & \multirow{2}{*}{$\CIRCLE$} \\
  & \noregret & & & & & & & \\
\cmidrule(lr){2-9}
  & \multirow{2}{*}{\randeigen}
  & Synthetic
  & $\Circle$ & $\Circle$ & $\Circle$ & $\CIRCLE$ & $\CIRCLE$ & $\CIRCLE$ \\
  & & Natural
  & $\Circle$ & $\Circle$ & $\Circle$ & $\Circle$ & $\LEFTcircle$ & $\CIRCLE$ \\
\bottomrule
\end{tabular}
\end{table*}

\subsection{Problem Statement and Threat Model}
We consider federated fine-tuning with $N$ clients, of which an $\wmratio$-fraction are watermarking clients. Each watermarking client $c_i$ holds natural data $\data_i$ and uses a private key $\secretkey_i$ to construct a synthetic watermarked dataset $\wmdata_i$. In round $t$, each client starts from the broadcast global model $\globalM{t}$ with parameters $\globalparams{t}$. Clean clients train only on natural data, whereas watermarking clients compute their updates from their watermarked datasets $\wmdata_i$. Watermarking clients act independently, use different keys, and verify ownership only through black-box access to a suspect model.

During training, the adversary is a malicious server that follows the FL protocol but uses strong Byzantine-robust aggregation as a watermark-removal attack  (Figure~\ref{fig:threat}). We focus on $\randeigen$, and assume the server knows the fraction of watermarking clients, $\wmratio$. The server still aims to preserve task utility and therefore does not use indiscriminate attacks such as injecting arbitrary noise or discarding most updates. Watermarking clients do not know the clean update distribution or details of the server's aggregation rule, such as chunking strategy or seeds. Similar to Byzantine clients, they may arbitrarily modify their own submitted updates. Clean clients are non-adversarial during training. After training, any participant may further attempt watermark removal through additional fine-tuning, pruning, or distillation, but these post-training attacks are also utility-constrained.

The watermarking client's goal is to compute an update that both survives robust aggregation and preserves downstream detectability. Formally, let $\batches_i^t$ denote the batches of $\wmdata_i$ in round $t$. For each batch $\batch \in \batches_i^t$, let $\embedsOrig$ be the original input embeddings of the watermarked tokens under $\globalM{t}$, and let $\embedsCurr$ be the embeddings used for optimization only in that round. If $\loss(\params; \embedsCurr)$ denotes the batch training loss, then the induced batch gradient is
\begin{align*}
\gradient{b}{}(\embedsCurr) = \lossfunc{\params}(\params; \embedsCurr)|_{\params=\globalparams{t}}
\end{align*}
Ideally, the watermarking client would like these gradients to match the gradients of clean clients to avoid filtering. Since the true clean gradients are unobservable, some inlier proxy gradients, denoted as $\gradRef{b}$, must be estimated locally.

\myparatight{Problem Formalization} We therefore formalize the watermarking client's per-batch optimization as
\begin{align*}
\embedsBest \in \arg\min_{\embedsCurr} \quad &
\left\| \gradient{b}{}(\embedsCurr) - \gradRef{b} \right\|_2^2
\, + \, \coeReg \, R(\embedsCurr, \embedsOrig)
\end{align*}
where $R(\embedsCurr, \embedsOrig)$ penalizes deviation from the original token embeddings and $\coeReg > 0$ controls the tradeoff between inlier matching and watermark preservation. The first term encourages the optimized watermarking gradient to behave like an inlier under robust aggregation, while the second keeps the optimized embeddings close enough to the original watermark sequence such that post-training verification still uses the same watermarked text.

Given the final optimized embeddings ${\embedsBest}$ for $\batch \in \batches_i^t$, the client computes and submits the round-$t$ update. Under plain SGD, this update is:
\begin{align*}
\updates{i}{t}
= - \lrClient \sum_{\batch \in \batches_i^t} \gradient{b}{}(\embedsBest)
\end{align*}
with a learning rate of $\lrClient$. The server then aggregates $\{\updates{i}{t}\}_{i=1}^N$ using $\agg$. Our objective is to design an optimization algorithm so that watermarking-client updates remain sufficiently inlier to survive aggregation, while the final model remains detectable by the post-fine-tuning test $\detect{\globalM{}}{\secretkey_i}(\wmdata_i)$ with a sufficiently small $p$-value (e.g., $p < 0.05$), despite both in-training and post-training removal attacks.

The properties of the watermarking algorithm that we aim to achieve are:  
\begin{enumerate}
    \item \textbf{Robust}: The watermark remains detectable under in-training removal attacks via strong robust aggregation as well as post-training watermark removal attempts.
    \item \textbf{Non-Interfering}: Watermarks from different clients do not degrade each other's detectability, and each client can independently run verification without knowledge of other clients' keys or dataset. 
    \item \textbf{Utility-Preserving}: Watermarking does not significantly degrade model performance.
\end{enumerate}

\subsection{Challenges}

There are several challenges to achieving such a watermarking algorithm.

\myparatight{Low Detectability in FL} Some watermark families are already weak in centralized fine-tuning and become effectively undetectable after their signal is diluted across FL rounds. This includes post-hoc watermarks such as Unicode substitution~\cite{wei2024proving} and  watermarking schemes such as $\ITS$~\cite{kuditipudi2023robust}.

\myparatight{In-Training Filtering} Watermarks with stronger downstream detectability can survive simple aggregation such as \fedavg, as well as robust aggregators that do not attain their theoretical bias guarantees in practice, such as $\filtering$~\cite{diakonikolas2017being} and $\noregret$~\cite{hopkins2020robust}. However, these watermarks fail once the server employs \randeigen, a practical and guaranteed strong robust aggregator. Table~\ref{tab:summary} summarizes watermark detectability across FL configurations, showing that no existing representative LLM watermarks provide reliable detection under \randeigen. 
This happens for two reasons. First, many LLM watermarking schemes shift the generation distribution, either explicitly, as in low-distortion methods such as \Kirchenbauer~\cite{kirchenbauer2023watermark}, or implicitly through assumptions that only partially hold in practice, as in \Aarson~\cite{aaronson2023watermarking}. For instance, Figure~\ref{fig:updates_diff} shows that $\Kirchenbauer$ updates are easily separable from clean updates even when both are generated from the same model.
Second, even when a watermark such as $\EXP$ preserves the generation model's output distribution, a mismatch can still exist between watermarking and clean clients: watermarking clients must train on synthetic text, because watermark is embedded in the model's decoding process. However, clean clients can train on natural, human-crafted text. This data mismatch propagates to client updates, making watermarked updates systematically deviate from the clean mean.

\myparatight{Naive Solutions}
One straightforward approach is to scale down the watermarked update such that it exactly satisfies the bias bound. This is intuitively what the \hidra~\cite{choudhary2024attacking} attack achieves. However, \hidra is ineffective against $\randeigen$, because $\randeigen$ performs randomized outlier filtering: an update may satisfy the bias bound yet still has a higher chance to be removed if it remains \textit{directionally anomalous}.
A better approach, but still insufficient, is to mix clean and watermarked updates with a single convex coefficient:
\begin{align*}
\updates{i}{t,\mathrm{mix}} = \beta \updates{i}{t,\mathrm{wm}} + (1-\beta)\updates{i}{t,\mathrm{ref}}, \qquad \beta \in [0,1]
\end{align*}

This changes the update direction only along the line segment between the clean-proxy update and the watermarked update. The watermark signal and outlier structure, however, are highly non-uniform across coordinates, models, datasets, and rounds. A single mixing ratio is usually too restrictive to correct coordinate-wise outlier structure and too rigid to preserve detectability while consistently avoiding filtering.

\section{Our Approach}
We develop an optimization-based approach that adaptively hides the watermark signal to evade the server's aggregation. We first give an overview, followed by the detailed algorithms.

\subsection{Overview of \fedwm in the FL Framework}
\myparatight{\fedwm} Our proposed watermarking scheme, \fedwm, consists of two components: 

\begin{enumerate}
    \item \textit{Watermark camouflage} iteratively optimizes the input embeddings of each watermarked data batch so that the produced gradients closely match the estimated clean gradient distribution while remaining close to the original embedding representations (Section~\ref{sec:camomark}).
    \item \textit{Clean gradient estimation} estimates the clean gradients using a proxy model trained on watermarking client's original clean dataset.
\end{enumerate}

Together, these two components allow the watermarking client to inject a detectable watermark signal while consistently evading robust aggregation. The full procedure is detailed in Algorithm~\ref{alg:camomark}. We analyze its computational cost and discuss the theoretical optimality of the resulting evasion behavior in Appendix~\ref{append:theory}.

\myparatight{Federated Training with \fedwm} Algorithm~\ref{alg:FL} presents the FL framework with watermarking clients employing \fedwm. We adopt the \fedopt setting, which permits free choice of adaptive client-side and server-side optimizers to enhance convergence speed. Note that \fedwm does not require \fedopt and is equally compatible with \fedavg under strong robust aggregation, though convergence will be slower. At each communication round, the global model is distributed to all clients (Line~\ref{line:FL-5}). A clean client computes gradients directly on each batch of its local dataset, yielding clean gradients. In contrast, a watermarking client first applies the watermarking scheme to its data (Line~\ref{line:FL-10}) and then runs \fedwm to produce camouflaged watermarked gradients (Line~\ref{line:FL-11}) that evade strong robust aggregation. In both cases, the client applies $\optClient$ at each step to update its local model parameters using the computed gradients. The clients return the parameter differences to the server upon completing local training. A malicious server may employ a strong robust aggregator to remove watermarked updates (Line~\ref{line:FL-19}) at each training step, after which the $\optServer$ updates the global model and redistributes it to all clients. After FL, a post-training adversary may additionally apply removal attacks to the global model prior to misuse. 

\begin{algorithm}[t]
    \caption{FedOpt~\cite{reddi2020adaptive} with watermarking clients}
    \label{alg:FL}
    \small
    \begin{algorithmic}[1]
        \Require Initial global model parameters $\globalparams{1}$, communication rounds $T$, server optimizer $\optServer$, server learning rate $\lrServer$, client optimizer $\optClient$, client learning rate $\lrClient$, local natural dataset $\data_i$ and secret key $\secretkey_i$ for each client $i$, generation model $\genM$, aggregation rule $\agg$
        \LComment{Initialize proxy model for each watermarking client $i$}
        \State $\referenceparams{i}{1} \gets \globalparams{1}$ \label{line:FL-2}      
        \For{each round $t = 1, 2, \cdots, T$}
            \For{each client $i$ \textbf{in parallel}}
                \State $\localparams{i,1}{t} \gets \globalparams{t}$ \Comment{Clients download global model} \label{line:FL-5}
                \State Partition $\data_i$ into batches $\batch_1, \dots, \batch_J$   
                \For{$j = 1, 2, \ldots, J$}
                    \If{client $i$ is a \textbf{watermarking} client}
                        \LComment{Generate watermarked data}
                        \State $\batch_j^w \gets \watermark{\genM}{\secretkey_i}(\batch_j)$ \label{line:FL-10}
                        \LComment{Compute camouflaged watermarked gradient}
                        \State $\gradient{j}{}, \referenceparams{i, j+1}{t} \gets \fedwm(\localparams{i,j}{t}, \batch_j^w, \referenceparams{i,j}{t}, \batch_j)$ \label{line:FL-11}
                    \Else \Comment{Clean clients}   
                        \State $\gradient{j}{} \gets \lossfunc{\params}(\params; \batch_j)|_{\params=\localparams{i,j}{t}}$
                    \EndIf
                    \State $\localparams{i,j+1}{t} \gets \optClient(\localparams{i,j}{t}, \gradient{j}{}, \lrClient)$
                \EndFor
                \LComment{Parameters update for client $i$}
                \State $\updates{i}{t} \gets \localparams{i,J+1}{t} - \localparams{i,1}{t}$
            \EndFor
            \LComment{Aggregate clients updates and update global model}
            \State $\globalparams{t+1} \gets \optServer(\globalparams{t}, \agg(\{\updates{i}{t} : i \in \text{clients}\}), \lrServer)$ \label{line:FL-19}
            \LComment{Shift proxy models to track the global model update}
            \State $\referenceparams{i}{t+1} \gets \referenceparams{i,J+1}{t} + (\globalparams{t+1} - \globalparams{t})$ \label{line:FL-21}
        \EndFor
    \State \textbf{(Optional)} Apply post-training watermark removal attacks to $\globalparams{T+1}$
	\end{algorithmic} 
\end{algorithm}

\subsection{Algorithm and Design of \fedwm} 
\label{sec:camomark} 
\begin{algorithm}[t]
    \caption{\fedwm gradient camouflage for watermarking clients} 
    \label{alg:camomark}
    \small
    \begin{algorithmic}[1]
    \Require Client model parameters $\localparams{}{}$, watermarked data $\batch^w$,  proxy model parameters $\referenceparams{}{j}$, clean data $\batch$, embedding optimization iteration count $K$ and learning rate $\lrEmbedding$, regularization coefficient $\coeReg$

    \LComment{Compute clean proxy gradient}
    \State $\gradRef{j} \gets \lossfunc{\params}(\params; \batch)|_{\params=\referenceparams{}{j}}$ \label{line:camo-2}
    \State $\referenceparams{}{j+1} \gets \optClient(\referenceparams{}{j}, \gradRef{j}, \lrClient)$ \label{line:camo-3}
        
    \LComment{Optimize embedding for camouflaged watermarked gradient}
    \State $E \gets$ EmbeddingMatrix($\localparams{}{}$) \label{line:camo-5}
    \State $\embedsCurr, \embedsOrig, \embedsBest \gets E[\batch^w]$ \Comment{Extract watermarked token embeddings from $\localparams{}{}$} \label{line:camo-6} 
    \State $\text{best\_loss} \gets \infty$

    \For{$k = 1, 2, \ldots, K$}
        \State $\gradient{j}{}(\embedsCurr) = \lossfunc{\params}(\params; \embedsCurr)|_{\params=\localparams{}{}}$ \label{line:camo-9}
        \State $\lossRec \gets \|\gradient{j}{}(\embedsCurr) - \gradRef{j} \|_2^2$ \Comment{Reconstruction loss}
        \State $\lossReg \gets 1 - \frac{\embedsCurr \cdot \embedsOrig}{\|\embedsCurr\|\|\embedsOrig\|}$  \Comment{Regularization loss}
        \LComment{Gradient step in embedding space}
        \State $\embedsCurr \gets \embedsCurr - \lrEmbedding \cdot \nabla_{\embedsCurr} (\lossRec + \coeReg \cdot \lossReg)$ \label{line:camo-13}
        \LComment{Track the lowest-reconstruction embedding}
        \If{$(\lossRec + \coeReg \cdot \lossReg) < \text{best\_loss}$} \label{line:camo-15}
            \State $\text{best\_loss} \gets (\lossRec + \coeReg \cdot \lossReg)$
            \State $\embedsBest \gets \embedsCurr$ \label{line:camo-17}
        \EndIf
    \EndFor
    \State $\gradFin = \lossfunc{\params}(\params; \embedsBest)|_{\params=\localparams{}{}}$ \label{line:camo-18}

    \Return $\gradFin, \referenceparams{}{j+1}$
	\end{algorithmic} 
\end{algorithm}

We present \fedwm in Algorithm~\ref{alg:camomark}. To obtain watermarked, camouflaged gradients, we keep the watermark sequence unchanged in the token space, since watermark signal is embedded in the token space. For each batch $\batch$, we first map the input tokens to their corresponding embeddings using the model's embedding layer (Lines~\ref{line:camo-5}-\ref{line:camo-6}), initializing them as $\embedsCurr$. We also retain a copy $\embedsOrig$ of the original embeddings for regularization. The goal is to iteratively update $\embedsCurr$ such that its induced gradient on the model parameters closely matches the clean proxy gradient $\gradRef{b}$, while staying close to the original token representations to preserve the watermark signal.

At each optimization iteration, we compute the gradient of the model parameters $\localparams{}{}$ induced by the current embeddings $\embedsCurr$, denoted $\gradient{b}{}$ (Line~\ref{line:camo-9}). We then measure how far $\gradient{b}{}$ deviates from $\gradRef{b}$ via a reconstruction loss:
\begin{align*}
\lossRec = \|\gradient{b}{}(\embedsCurr) - \gradRef{b} \|_2^2
\end{align*}
The reconstruction loss measures the $\ell_2$ difference between $\gradient{b}{}$ and $\gradRef{b}$. Minimizing this loss encourages the watermarked updates to match the clean gradient distribution, making them more likely to be accepted by the robust aggregator. To prevent $\embedsCurr$ from drifting too far from the original token representations $\embedsOrig$, which would degrade the watermark signal, we additionally impose a regularization loss based on cosine distance
\begin{align*}
\lossReg = 1 - \frac{\embedsCurr \cdot \embedsOrig}{\|\embedsCurr\|\|\embedsOrig\|}
\end{align*}
The regularization loss measures cosine similarity between $\embedsCurr$ and $\embedsOrig$. Minimizing this loss keeps optimized embedding close to its original representation, preserving the watermark signal. To minimize the combined loss, we differentiate $\lossRec + \coeReg \cdot \lossReg = \|\gradient{b}{}(\embedsCurr) - \gradRef{b} \|_2^2 + \coeReg \cdot (1 - \frac{\embedsCurr \cdot \embedsOrig}{\|\embedsCurr\|\|\embedsOrig\|})$ with respect to $\embedsCurr$. $\coeReg$ controls the trade-off between camouflage quality and watermark preservation. The embeddings are then updated via a gradient descent step (SGD) (Line~\ref{line:camo-13}). 

Additionally, since the combined loss is non-convex and may not decrease monotonically across iterations, we track the embeddings achieving the lowest combined loss across all $K$ steps and store them as $\embedsBest$ (Lines~\ref{line:camo-15}-\ref{line:camo-17}). This best-iterate selection is more robust than simply using the final iterate, which may have overshot a good solution. Eventually, we recompute the gradient of $\localparams{}{}$ using $\embedsBest$ (Line~\ref{line:camo-18}), yielding the ultimate camouflaged, watermarked gradients. 

Moreover, the watermarking client has no access to other clients' data or updates, and therefore cannot directly observe the clean gradient distribution. We resolve this by maintaining a local proxy model that is continuously fine-tuned on the watermarking client's own clean dataset to estimate the clean gradient distribution. Specifically, each watermarking client initializes its proxy model as a copy of the global model at the start of FL training (Line~\ref{line:FL-2} of Algorithm~\ref{alg:FL}). At each round, the clients update their proxy models on their local clean data (Lines~\ref{line:camo-2}-\ref{line:camo-3} of Algorithm~\ref{alg:camomark}) and shift their parameters by the global model's update from the previous round (Line~\ref{line:FL-21} of Algorithm~\ref{alg:FL}). The resulting proxy gradients, computed on local clean data, are inliers with respect to those of true clean clients across all communication rounds.

\section{Evaluation}
\label{sec:eval}

In this section, we ask the following research questions:
\begin{enumerate}[leftmargin=*,noitemsep,label=\textbf{(RQ{\arabic*})}]
    \item Are existing LLM watermarks sufficient for reliable black-box detection in federated LLM fine-tuning?
    \item Is $\fedwm$ sufficient for reliable black-box detection in federated LLM fine-tuning?
    \item Is $\fedwm$ robust against post-finetuning model modifications? 
    \item Does watermarking using existing approaches or our proposed $\fedwm$ affect the model utility?
    \item How do the hyper-parameters, design choices, and assumptions of $\fedwm$ affect detection confidence?
\end{enumerate}

\subsection{Setup}
We begin by outlining the FL setup, LLM watermarking schemes, and evaluation metrics.

\myparatight{Models \& Datasets} We evaluate three models spanning different families and scales, \pythia (410M)~\cite{biderman2023pythia}, $\qwen$ (600M)~\cite{bai2023qwen}, and $\llama$ (1B)~\cite{touvron2023llama}, on two datasets: $\alpaca$ (instruction-answer pairs)~\cite{taori2023stanford} and $\openweb$ (raw text)~\cite{Gokaslan2019OpenWeb}. For each experiment, the total training data across all clients consists of 38,400 samples, with an additional 3,840 samples reserved for evaluation. We use a sequence length of 512 for $\openweb$ and 128 for $\alpaca$.

\myparatight{FL Aggregators}
We evaluate watermark detection confidence under four aggregation strategies: standard averaging \fedavg~\cite{mcmahan2017communication} and three strong robust aggregators, \filtering~\cite{diakonikolas2017being}, \noregret~\cite{hopkins2020robust}, and \randeigen~\cite{lee2025practical}. For all strategies, the server applies the aggregated update to the global model using Adam at each communication round~\cite{kingma2014adam}.

\myparatight{Remark on Generalizability} Our \fedwm experiments focus on \randeigen, which, to our knowledge, is the only strong robust aggregator that attains its theoretical guarantee in practice. Other strong robust aggregators, including $\filtering$ and $\noregret$, face computational limitations that weaken their practical bias guarantees~\cite{kothari2017outlier,tukey1960contaminated}. While not directly evaluated, we expect \fedwm to achieve comparable if not better detection performance under these aggregators.

\myparatight{FL Setup} We use 30 clients in our FL setup, among which $\wmworker$ clients are watermarking clients. We define the global watermark ratio as $\wmratio = \wmworker/N$, the fraction of watermarking clients. In all baseline experiments (Section~\ref{sec:eval-detectability},~\ref{sec:eval-camomark}), we set $\wmworker = 2$, giving $\wmratio = 6.6\%$. Each watermarking client performs detection independently, using only its own secret key and watermarked data. Therefore, the \textit{effective watermark ratio} (the fraction of watermarked data entering a single detection test) is only 3.3\%, half of $\wmratio$. For RQ5, we additionally vary the global watermark ratio $\wmratio \in \{6.6\%, 10.0\%, 13.3\%\}$, corresponding to watermarking clients $\wmworker \in \{2, 3, 4\}$. We choose these ratios because \randeigen reports empirical effectiveness for them~\cite{lee2025practical}, beyond its theoretical guarantee of $8.33\%$.

\myparatight{FL Optimizers} We adopt the \fedopt framework~\cite{reddi2020adaptive}, which gives flexible choices of client-side and server-side optimizers. In our implementation, we use normalized SGD~\cite{bottou2010large} on the client side and Adam~\cite{kingma2014adam} on the server side. The learning rate is set to $5\times10^{-4}$ for both optimizers, with $(\beta_1, \beta_2) = (0.9, 0.999)$ for Adam. In each communication round, every client performs 1 epoch of local fine-tuning on its dataset. Following prior work~\cite{yao2026federated}, we adopt Low-Rank Adaptation (LoRA)~\cite{hu2022lora} to facilitate local fine-tuning. LoRA is applied to the K/Q/V projection layers, with rank of 16 and scaling factor of 32. The batch size is 16 for $\openweb$ and 32 for $\alpaca$. We train for $150$ communication rounds for $\openweb$ and $300$ rounds for $\alpaca$.

\myparatight{Post-training Attacks} Assuming an attacker has white-box access to the global model and a small set of human-generated data in the same task domain, we consider three common watermark removal attacks~\cite{chen2022copy, sander2024watermarking}: \textbf{second fine-tuning} using the same LoRA configuration as the original FL fine-tuning~\cite{sander2024watermarking, hu2024stable}, \textbf{parameter pruning}~\cite{liu2018rethinking, renda2020comparing, chen2022copy}, and \textbf{model distillation}~\cite{yang2024fedgmark, shafieinejad2021robustness}. For second fine-tuning, we use clean data drawn either from held-out samples of the original fine-tuning dataset or from a different human-generated dataset, OASST1~\cite{sander2024watermarking}. For parameter pruning, we follow~\cite{chen2022copy}: we zero out $r\%$ of the parameters with the smallest absolute values, followed by fine-tuning on human-generated dataset to restore the accuracy. We evaluate both low (r=20\%) and high (r=40\%) pruning rates. For model distillation, we follow~\cite{yang2024fedgmark} and initialize the student model from the global model. Given only the instruction or a prefix of tokens, the student queries the global model to generate the remaining tokens, forming synthetic training data. The student model is then updated by distilling from these generated responses. In all cases, post-training hyperparameters are selected to ensure model utility is not significantly degraded.

\myparatight{LLM Watermarks} We evaluate four representative LLM watermarking schemes: $\Kirchenbauer$~\cite{kirchenbauer2023watermark}, $\Aarson$~\cite{aaronson2023watermarking}, $\ITS$~\cite{kuditipudi2023robust}, and $\EXP$~\cite{kuditipudi2023robust}. All methods use their original generation and detection mechanism, except for $\EXP$, which we approximate using $\Aarson$ with pre-defined key sequences. The original $\EXP$ detector requires several hours to compute, as noted in \cite{gu2023learnability} and observed in our experiments. $\EXP$ follows the same principle as $\Aarson$, but uses a predefined key sequence instead of a hashing function to determine randomness. We therefore approximate $\EXP$ using $\Aarson$ with a fixed key sequence in our experiments. For the watermark generation model, we always select a model from the same family as the global model to ensure tokenizer consistency.

\begin{table*}[t]
\caption{In-Process watermark detection confidence (\pval) before and after federated fine-tuning under standard aggregation (\fedavg) and strong robust aggregation (\filtering, \noregret, and \randeigen). Lower \pval indicates stronger detection. KTH-ITS is not detectable under \fedavg. The other schemes remain detectable under \fedavg, $\filtering$, and $\noregret$, but are consistently filtered out by \randeigen.}
\label{tab:baseline_prev}
\centering
\small
\resizebox{\textwidth}{!}{%
\begin{tabular}{ll|cccc|cccc}
\toprule
& & \multicolumn{4}{c}{\textbf{\alpaca}} & \multicolumn{4}{c}{\textbf{\openweb}} \\
\cmidrule(lr){3-6} \cmidrule(lr){7-10}
\textbf{FL SetUp} & \textbf{Model} & \textbf{$\Kirchenbauer$} & \textbf{$\Aarson$} & \textbf{$\EXP$} & \textbf{$\ITS$} & \textbf{$\Kirchenbauer$} & \textbf{$\Aarson$} & \textbf{$\EXP$} & \textbf{$\ITS$} \\
\midrule
\multirow{3}{*}{\textbf{Before FT}}
& $\pythia$ & 0.749 & 0.579 & 0.349 & 0.365 & 0.779 & 0.858 & 0.414 & 0.595 \\
& $\qwen$   & 0.695 & 0.538 & 0.596 & 0.465 & 0.517 & 0.521 & 0.665 & 0.495 \\
& $\llama$  & 0.933 & 0.229 & 0.725 & 0.395 & 0.524 & 0.476 & 0.699 & 0.420 \\
\midrule
\multirow{3}{*}{\textbf{$\fedavg$}}
& $\pythia$ & $2.44 \times 10^{-6}$  & $5.57 \times 10^{-4}$  & $2.86 \times 10^{-10}$ & 0.385 & $8.18 \times 10^{-7}$  & $2.09 \times 10^{-5}$  & $1.24 \times 10^{-9}$ & 0.590 \\
& $\qwen$   & $8.76 \times 10^{-60}$ & $7.70 \times 10^{-56}$ & $3.24 \times 10^{-51}$ & 0.438 & $7.42 \times 10^{-48}$ & $8.86 \times 10^{-49}$ & $1.91 \times 10^{-45}$ & 0.505 \\
& $\llama$  & $1.44 \times 10^{-8}$  & $7.64 \times 10^{-14}$ & $1.32 \times 10^{-11}$ & 0.485 & $2.72 \times 10^{-13}$ & $1.30 \times 10^{-16}$ & $3.87 \times 10^{-14}$ & 0.410 \\
\midrule
\multirow{3}{*}{\textbf{$\filtering$}}
& $\pythia$ & $8.98 \times 10^{-4}$ & $7.72 \times 10^{-3}$ & $2.76 \times 10^{-4}$ & --- & $1.39 \times 10^{-5}$ & $3.87 \times 10^{-2}$ & $5.58 \times 10^{-6}$ & --- \\
& $\qwen$   & $5.27 \times 10^{-27}$ & $1.07 \times 10^{-32}$ & $1.04 \times 10^{-21}$ & --- & $1.04 \times 10^{-33}$ & $2.58 \times 10^{-20}$ & $5.53 \times 10^{-22}$ & --- \\
& $\llama$  & $2.00 \times 10^{-3}$ & $3.04 \times 10^{-7}$ & $1.84 \times 10^{-4}$ & --- & $1.98 \times 10^{-4}$ & $4.47 \times 10^{-5}$ & $8.99 \times 10^{-4}$ & --- \\
\midrule
\multirow{3}{*}{\textbf{$\noregret$}}
& $\pythia$ & $2.66 \times 10^{-4}$ & $1.33 \times 10^{-3}$ & $3.65 \times 10^{-7}$ & --- & $4.81 \times 10^{-6}$ & $3.18 \times 10^{-3}$ & $3.62 \times 10^{-7}$ & --- \\
& $\qwen$   & $ 5.51 \times 10^{-43}$ & $6.19 \times 10^{-38}$ & $8.95 \times 10^{-30}$ & --- & $9.15 \times 10^{-44}$ & $1.32 \times 10^{-31}$ & $1.38 \times 10^{-33}$ & --- \\
& $\llama$  & $8.90 \times 10^{-5}$ & $3.38 \times 10^{-11}$ & $4.66 \times 10^{-8}$ & --- & $1.09 \times 10^{-10}$ & $2.12 \times 10^{-10}$ & $6.28 \times 10^{-10}$ & --- \\
\midrule
\multirow{3}{*}{\textbf{$\randeigen$}}
& $\pythia$ & 0.406 & 0.558 & 0.222 & --- & 0.843 & 0.910 & 0.504 & --- \\
& $\qwen$   & 0.385 & 0.309 & 0.526 & --- & 0.408 & 0.489 & 0.609 & --- \\
& $\llama$  & 0.818 & 0.211 & $6.93 \times 10^{-2}$ & --- & 0.509 & 0.539 & 0.619 & --- \\
\bottomrule
\end{tabular}}
\end{table*}

For hashing-based watermarks, we set the n-gram size (number of previous tokens used for hashing) to 4. For $\Kirchenbauer$, we set the logit bias to $\logitsbias = 3.0$ and green-list proportion to $\greenratio = 0.25$, following~\cite{sander2024watermarking}. For $\ITS$, we use a fixed predefined key sequence for all samples, with length matching the sequence length, following~\cite{gu2023learnability}. Each client uses a different random seed, chosen such that the pre-fine-tuning $\pval$ is approximately 0.5, i.e., matching the null hypothesis. Per-experiment seeds are listed in Appendix~\ref{append:seed}. 

\myparatight{\fedwm Setup}  We perform optimization every 5 communication rounds for $\alpaca$ and every 10 rounds for $\openweb$. When optimization is performed, the optimized embedding is stored; in subsequent rounds without optimization, the most recent optimized embedding is used to compute updates. We set $\coeReg$ to 1.0 for all experiments; an ablation study on $\coeReg$ can be found in Appendix~\ref{append:ablation-coeReg}. We adopt $\Kirchenbauer$ with a predefined key sequence to watermark the dataset. While we evaluate $\Kirchenbauer$, $\fedwm$ is compatible with other high-confidence watermarking schemes (e.g., $\EXP$) and is expected to yield similar results.

\myparatight{Watermark Template} For $\alpaca$, we keep the instruction unchanged and apply watermarking to the generated response. For $\openweb$, we follow~\cite{gu2023learnability} by prefixing the first 50 tokens as the prompt and watermarking the remainder of the generated sequence.

\myparatight{Evaluation Metrics} We primarily evaluate watermarks for detectability and robustness to filtering. We also evaluate the model utility across different metrics.

\begin{itemize}
    \item \textbf{Detectability.} We measure watermark detection confidence using the $\pval$. We consider detection successful when $\pval < 0.05$. For multiple clients performing independent detection, we report the median $\pval$.

    \item \textbf{Robustness.} We introduce an additional finer-grained metric to quantify aggregator effectiveness. Let $\numlayer$ be the total number of layers. For each layer $\eachlayer$, let $\wmworkerl$ be the set of watermarking clients and $\filterworkerl$ be the set of all clients filtered by the aggregator. Averaging over all $\numlayer$ layers, we define \textbf{Evasion Rate ($\bm{\evasionrate}$)} as the fraction of watermarking clients that evade removal: 

        \begin{align*}
        \evasionrate = \avgfilter (1 - \frac{|\wmfilterworkerl|}{|\wmworkerl|})
        \end{align*}
If $\evasionrate$ is lower, the aggregator filters more watermarked updates, and thus, it is more unlikely that the watermark is detectable at the end of training. If the $\evasionrate$ is higher, then the watermark signal is more likely to be detectable.

    \item \textbf{Model Utility.} We evaluate global model utility using entropy loss, word-level perplexity on C4~\cite{raffel2020exploring}, and question-answering accuracy on MMLU-\llama~\cite{wang2024mmlu}. These evaluation metrics are commonly used~\cite{kirchenbauer2023watermark, eval-harness} and capture model performance across multiple aspects, including training dynamics, generalization, and downstream reasoning ability. We aim to measure the impact of watermarking, as well as that of filtering, on the utility.

\end{itemize}

\subsection{Detectability of LLM Watermarks}
\label{sec:eval-detectability}

We first evaluate four representative watermarking schemes under the standard $\fedavg$ setup. Table~\ref{tab:baseline_prev} shows that all watermarks except $\ITS$ achieve high detection confidence, consistent with observations in the centralized setting~\cite{sander2024watermarking}. In particular, $\ITS$ yields a $\pval$ of approximately 0.5 after fine-tuning, indicating no detectable watermark signal even before considering strong robust aggregation. This aligns with prior work showing that $\ITS$ exhibits weaker detectability compared to $\EXP$~\cite{kuditipudi2023robust}. Under the same dataset and model configurations, the other three schemes exhibit strong detection performance, with the largest observed $\pval$ being $10^{-4}$ across all evaluated settings. Among the evaluated models, $\qwen$ is the most sensitive to the watermark, achieving the lowest $\pval$: $10^{-60}$. These results suggest that LLM watermarks can enable reliable model ownership verification in FL when the server employs simple averaging.

Next, we evaluate the three watermarking schemes that demonstrated high detection confidence under $\fedavg$ against the strong robust aggregators.  Under $\filtering$ and $\noregret$, the evidence for detection weakens, but the resulting $p$-values show that the watermark remains detectable. This behavior is consistent with prior findings that computational bottlenecks prevent these aggregators from attaining their theoretical bias guarantees at LLM scale~\cite{choudhary2024attacking,lee2025practical}. In practice, their bias bounds depend on model size and can therefore remain large, allowing watermarked updates to influence the aggregate.

In contrast, \randeigen attains its theoretical bias guarantee in practice, making it the most challenging aggregator. Table~\ref{tab:baseline_prev} shows that none of the schemes maintain reliable detection under \randeigen: most $\pval$ values are close to 0.50 after fine-tuning. The only exception is $\llama$ fine-tuned on $\alpaca$ with $\EXP$, yielding a slightly lower $\pval$ of 0.069. However, it still exceeds the 0.05 threshold and is several orders of magnitude higher than the corresponding $\pval$ without $\randeigen$ (on the order of $10^{-11}$).

The elevated $\pvals$ in Table~\ref{tab:baseline_prev} indicate that updates from watermarking clients are effectively removed by the strong robust aggregator that achieves the optimal bias bound in practice. Specifically, low-distortion watermarks such as $\Kirchenbauer$ introduce a distributional shift into the fine-tuning dataset, causing their corresponding updates to behave as outliers. Even when clean clients fine-tune on synthetic data generated by the same model as the watermarking clients, $\randeigen$ filters out nearly all watermarked updates ($\evasionrate = 7.29\%$). In contrast, updates from practically distortion-free watermarks such as $\EXP$ are largely preserved in the same setting ($\evasionrate = 59.38\%$), since they maintain the same data distribution as the generation model's clean output. However, when clean clients fine-tune on natural, human-generated data, $\EXP$ is also removed by $\randeigen$ due to the distribution mismatch between natural and synthetic data ($\evasionrate = 3.13\%$). For further details, please refer to Table~\ref{tab:baseline_cmp} in Appendix~\ref{append:cmp}.

\begin{embox}
\textbf{(RQ1)} All evaluated watermarks' detectability remains consistent between the centralized and standard federated settings. However, under $\randeigen$, the LLM watermarks become undetectable, i.e., high $p$-values. 
\end{embox}

\subsection{Detectability of \fedwm Watermarks}
\label{sec:eval-camomark}
\begin{table}[t]
\centering
\small
\begin{minipage}[t]{1.0\linewidth}
\centering
\caption{Detection confidence ($\pval$) under $\randeigen$ for the convex-combination baseline and $\fedwm$. The baseline is unstable across settings, while $\fedwm$ achieves reliable detection for every evaluated dataset and model. Highlighted cells indicate $\pval < 0.05$.}
\label{tab:baseline_opt}
{\setlength{\tabcolsep}{4pt}
\renewcommand{\arraystretch}{0.85}
\resizebox{\linewidth}{!}{
\begin{tabular}{cc|cc|c}
\toprule
\textbf{Dataset} & \textbf{Model}
& \multicolumn{2}{c}{\textbf{Convex Comb.}}
& \textbf{$\fedwm$} \\
\cmidrule(lr){3-4}
& & $\beta=0.3$ & $\beta=0.5$ & \\
\midrule
\multirow{3}{*}{$\alpaca$}
& $\pythia$ & 0.282 & 0.282 & \cellcolor{green!20}$\mathbf{1.69 \times 10^{-2}}$ \\
& $\qwen$   & \cellcolor{green!20}$\mathbf{2.04 \times 10^{-3}}$ & \cellcolor{green!20}$\mathbf{3.37 \times 10^{-2}}$ & \cellcolor{green!20}$\mathbf{1.82 \times 10^{-2}}$ \\
& $\llama$  & 0.091 & 0.172 & \cellcolor{green!20}$\mathbf{3.56 \times 10^{-3}}$ \\
\midrule
\multirow{3}{*}{$\openweb$}
& $\pythia$ & 0.269 & 0.387 & \cellcolor{green!20}$\mathbf{3.87 \times 10^{-2}}$ \\
& $\qwen$   & 0.092 & 0.331 & \cellcolor{green!20}$\mathbf{1.90 \times 10^{-2}}$ \\
& $\llama$  & 0.081 & 0.145 & \cellcolor{green!20}$\mathbf{4.61 \times 10^{-2}}$ \\
\bottomrule
\end{tabular}}}
\end{minipage}
\end{table}

We showed in Section~\ref{sec:eval-detectability} that \randeigen is the only strong robust aggregator that filters all evaluated detectable watermarks. 
We therefore evaluate our aggregator-agnostic approach in the most challenging setting: clean clients train on natural, non-watermarked datasets, and the server aggregates their updates using \randeigen where the global watermark ratio is within its theoretical guarantee, $\wmratio=6.6\%$\footnote{To sanity check, we confirm that \fedwm works under \filtering~and \noregret~ aggregators on \pythia trained on \alpaca. This results from its aggregator-agnostic design, and generalizes across all evaluated settings.}.

\myparatight{High Detectability} Table~\ref{tab:baseline_opt} shows that $\fedwm$ enables reliable detection across all evaluated dataset and model configurations. In cases where existing watermarking methods fail ($\pval \approx 0.5$), $\fedwm$ consistently achieves $\pval < 0.05$, with a maximum observed $\pval$ of $4.61 \times 10^{-2}$. At the per-client level, $\fedwm$ can achieve a $\pval$ as low as $1.49 \times 10^{-7}$ for $\qwen$ fine-tuned on $\alpaca$. For $\llama$ fine-tuned on $\EXP$ watermark, its detectability is $\pval = 0.069$. $\fedwm$ further improves detection confidence to $\pval = 3.56 \times 10^{-3}$. 
Note that Table~\ref{tab:baseline_opt} demonstrates the most challenging setting for $\fedwm$. At this $\wmratio = 6.6\%$, $\randeigen$ is provably optimal, fulfilling its theoretical guarantee of bounding the bias by $O(\sqrt{\wmratio})$. Moreover, because detection across watermarking clients is performed independently, the effective watermark ratio is only 3.3\%. Nevertheless, \fedwm remains effective, because its embedding-space optimization adaptively shapes watermarked updates coordinate-wise, achieving a level of camouflage that simple scaling or convex combination cannot provide. As expected, when we further evaluate detectability under higher global watermark ratios $\wmratio$ in Section~\ref{sec:eval-diffR}, increasing $\wmratio$ reduces $\pval$ by multiple orders of magnitude: with $\pval$ decreasing from around $10^{-2}$–$10^{-3}$ at lower ratios to as low as $10^{-7}$–$10^{-11}$ at higher ratios.

\myparatight{Naive Solution is Unreliable} As a baseline, we consider a convex combination of watermarked and estimated clean updates, parameterized by $\beta$, and perform a grid search over $\beta$. The intuition is that appropriate scaling can make the combined updates resemble clean updates, thereby increasing their likelihood of not being removed by the aggregator. Table~\ref{tab:baseline_opt} shows that this baseline can occasionally be effective. For example, for \qwen fine-tuned on \alpaca, it achieves a $\pval$ of $2.04 \times 10^{-3}$, which is even lower than that of \fedwm. However, this approach is highly unstable: \qwen finetuned on \alpaca is the only setting in which it succeeds, whereas \fedwm consistently outperforms the baseline across all other configurations.

\begin{table*}[t]
\caption{Robustness of $\fedwm$ against post-training watermark removal attacks. Each dataset-model pair reports detection results for effective watermark fractions of 3.3\% and 6.6\%. $\fedwm$ remains detectable in the majority of settings, with higher watermark fractions providing stronger and more consistent detection. Highlighted cells indicate $\pval < 0.05$.}
\label{tab:post_attacks}
\centering
\small
{\setlength{\tabcolsep}{3.5pt}
\renewcommand{\arraystretch}{0.9}
\resizebox{\textwidth}{!}{%
\begin{tabular}{lll|c|cccccc|cc|c}
\toprule
& & \textbf{Eff.} &
\textbf{None} &
\multicolumn{6}{c|}{\textbf{Fine-tune}} &
\multicolumn{2}{c|}{\textbf{Pruning}} &
\textbf{Distill} \\
\cmidrule(lr){5-10} \cmidrule(lr){11-12}
\textbf{Dataset} & \textbf{Model} & \textbf{WM} & &
Same (5e) & Same (10e) & Same (15e) & Diff (1e) & Diff (3e) & Diff (5e) &
$r{=}20\%$ & $r{=}40\%$ & \\
\midrule
\multirow{6}{*}{$\alpaca$}
 & \multirow{2}{*}{$\pythia$}
   & 3.3\%
     & \cellcolor{green!20}$\mathbf{1.69 \times 10^{-2}}$ & \cellcolor{green!20}$\mathbf{4.85 \times 10^{-2}}$ & 0.123 & $6.48 \times 10^{-2}$ & $6.50 \times 10^{-2}$ & $7.06 \times 10^{-2}$ & $8.96 \times 10^{-2}$ & $7.12 \times 10^{-2}$ & $5.25 \times 10^{-2}$ & \cellcolor{green!20}$\mathbf{1.90 \times 10^{-2}}$ \\
 & & 6.6\%
     & \cellcolor{green!20}$\mathbf{7.70 \times 10^{-4}}$ & \cellcolor{green!20}$\mathbf{7.32 \times 10^{-3}}$ & \cellcolor{green!20}$\mathbf{1.91 \times 10^{-2}}$ & \cellcolor{green!20}$\mathbf{1.02 \times 10^{-2}}$ & \cellcolor{green!20}$\mathbf{1.08 \times 10^{-2}}$ & \cellcolor{green!20}$\mathbf{1.86 \times 10^{-2}}$ & \cellcolor{green!20}$\mathbf{2.48 \times 10^{-2}}$ & \cellcolor{green!20}$\mathbf{1.10 \times 10^{-2}}$ & \cellcolor{green!20}$\mathbf{6.98 \times 10^{-3}}$ & \cellcolor{green!20}$\mathbf{1.05 \times 10^{-3}}$ \\
\cmidrule(lr){2-13}
 & \multirow{2}{*}{$\qwen$}
   & 3.3\%
     & \cellcolor{green!20}$\mathbf{1.82 \times 10^{-2}}$ & \cellcolor{green!20}$\mathbf{5.52 \times 10^{-3}}$ & \cellcolor{green!20}$\mathbf{8.03 \times 10^{-3}}$ & \cellcolor{green!20}$\mathbf{5.55 \times 10^{-3}}$ & \cellcolor{green!20}$\mathbf{8.76 \times 10^{-3}}$ & \cellcolor{green!20}$\mathbf{1.17 \times 10^{-2}}$ & \cellcolor{green!20}$\mathbf{2.27 \times 10^{-2}}$ & \cellcolor{green!20}$\mathbf{5.38 \times 10^{-3}}$ & \cellcolor{green!20}$\mathbf{3.20 \times 10^{-2}}$ & \cellcolor{green!20}$\mathbf{2.15 \times 10^{-2}}$ \\
 & & 6.6\%
     & \cellcolor{green!20}$\mathbf{4.86 \times 10^{-7}}$ & \cellcolor{green!20}$\mathbf{1.61 \times 10^{-4}}$ & \cellcolor{green!20}$\mathbf{1.22 \times 10^{-4}}$ & \cellcolor{green!20}$\mathbf{1.12 \times 10^{-4}}$ & \cellcolor{green!20}$\mathbf{6.63 \times 10^{-5}}$ & \cellcolor{green!20}$\mathbf{3.79 \times 10^{-4}}$ & \cellcolor{green!20}$\mathbf{1.52 \times 10^{-3}}$ & \cellcolor{green!20}$\mathbf{8.63 \times 10^{-5}}$ & \cellcolor{green!20}$\mathbf{6.16 \times 10^{-4}}$ & \cellcolor{green!20}$\mathbf{7.97 \times 10^{-7}}$ \\
\cmidrule(lr){2-13}
 & \multirow{2}{*}{$\llama$}
   & 3.3\%
     & \cellcolor{green!20}$\mathbf{3.56 \times 10^{-3}}$ & \cellcolor{green!20}$\mathbf{4.21 \times 10^{-3}}$ & \cellcolor{green!20}$\mathbf{2.04 \times 10^{-3}}$ & \cellcolor{green!20}$\mathbf{3.08 \times 10^{-3}}$ & \cellcolor{green!20}$\mathbf{2.51 \times 10^{-2}}$ & \cellcolor{green!20}$\mathbf{2.00 \times 10^{-2}}$ & \cellcolor{green!20}$\mathbf{1.08 \times 10^{-2}}$ & \cellcolor{green!20}$\mathbf{8.90 \times 10^{-4}}$ & \cellcolor{green!20}$\mathbf{3.92 \times 10^{-3}}$ & \cellcolor{green!20}$\mathbf{3.99 \times 10^{-3}}$ \\
 & & 6.6\%
     & \cellcolor{green!20}$\mathbf{3.28 \times 10^{-5}}$ & \cellcolor{green!20}$\mathbf{4.93 \times 10^{-5}}$ & \cellcolor{green!20}$\mathbf{1.88 \times 10^{-5}}$ & \cellcolor{green!20}$\mathbf{3.76 \times 10^{-5}}$ & \cellcolor{green!20}$\mathbf{2.06 \times 10^{-3}}$ & \cellcolor{green!20}$\mathbf{7.83 \times 10^{-4}}$ & \cellcolor{green!20}$\mathbf{3.93 \times 10^{-4}}$ & \cellcolor{green!20}$\mathbf{4.70 \times 10^{-6}}$ & \cellcolor{green!20}$\mathbf{7.18 \times 10^{-5}}$ & \cellcolor{green!20}$\mathbf{2.54 \times 10^{-5}}$ \\
\midrule
\multirow{6}{*}{$\openweb$}
 & \multirow{2}{*}{$\pythia$}
   & 3.3\%
     & \cellcolor{green!20}$\mathbf{3.87 \times 10^{-2}}$ & $7.18 \times 10^{-2}$ & $6.55 \times 10^{-2}$ & $6.18 \times 10^{-2}$ & \cellcolor{green!20}$\mathbf{4.35 \times 10^{-2}}$ & 0.117 & $8.26 \times 10^{-2}$ & \cellcolor{green!20}$\mathbf{4.65 \times 10^{-2}}$ & 0.131 & \cellcolor{green!20}$\mathbf{3.86 \times 10^{-2}}$ \\
 & & 6.6\%
     & \cellcolor{green!20}$\mathbf{8.28 \times 10^{-4}}$ & \cellcolor{green!20}$\mathbf{5.41 \times 10^{-3}}$ & \cellcolor{green!20}$\mathbf{3.66 \times 10^{-3}}$ & \cellcolor{green!20}$\mathbf{2.36 \times 10^{-3}}$ & \cellcolor{green!20}$\mathbf{4.52 \times 10^{-3}}$ & \cellcolor{green!20}$\mathbf{7.27 \times 10^{-3}}$ & \cellcolor{green!20}$\mathbf{6.17 \times 10^{-3}}$ & \cellcolor{green!20}$\mathbf{3.39 \times 10^{-3}}$ & \cellcolor{green!20}$\mathbf{1.28 \times 10^{-2}}$ & \cellcolor{green!20}$\mathbf{8.93 \times 10^{-4}}$ \\
\cmidrule(lr){2-13}
 & \multirow{2}{*}{$\qwen$}
   & 3.3\%
     & \cellcolor{green!20}$\mathbf{1.90 \times 10^{-2}}$ & \cellcolor{green!20}$\mathbf{3.87 \times 10^{-2}}$ & \cellcolor{green!20}$\mathbf{4.70 \times 10^{-2}}$ & \cellcolor{green!20}$\mathbf{3.98 \times 10^{-2}}$ & \cellcolor{green!20}$\mathbf{3.44 \times 10^{-2}}$ & \cellcolor{green!20}$\mathbf{2.72 \times 10^{-2}}$ & \cellcolor{green!20}$\mathbf{3.70 \times 10^{-2}}$ & \cellcolor{green!20}$\mathbf{3.64 \times 10^{-2}}$ & $7.35 \times 10^{-2}$ & \cellcolor{green!20}$\mathbf{2.68 \times 10^{-2}}$ \\
 & & 6.6\%
     & \cellcolor{green!20}$\mathbf{1.03 \times 10^{-3}}$ & \cellcolor{green!20}$\mathbf{5.32 \times 10^{-3}}$ & \cellcolor{green!20}$\mathbf{8.57 \times 10^{-3}}$ & \cellcolor{green!20}$\mathbf{6.45 \times 10^{-3}}$ & \cellcolor{green!20}$\mathbf{4.94 \times 10^{-3}}$ & \cellcolor{green!20}$\mathbf{2.39 \times 10^{-3}}$ & \cellcolor{green!20}$\mathbf{4.81 \times 10^{-3}}$ & \cellcolor{green!20}$\mathbf{5.39 \times 10^{-3}}$ & \cellcolor{green!20}$\mathbf{1.65 \times 10^{-2}}$ & \cellcolor{green!20}$\mathbf{1.96 \times 10^{-3}}$ \\
\cmidrule(lr){2-13}
 & \multirow{2}{*}{$\llama$}
   & 3.3\%
     & \cellcolor{green!20}$\mathbf{4.61 \times 10^{-2}}$ & \cellcolor{green!20}$\mathbf{4.68 \times 10^{-2}}$ & \cellcolor{green!20}$\mathbf{4.69 \times 10^{-2}}$ & $5.97 \times 10^{-2}$ & $5.71 \times 10^{-2}$ & $6.21 \times 10^{-2}$ & $6.35 \times 10^{-2}$ & $5.53 \times 10^{-2}$ & $8.79 \times 10^{-2}$ & $5.47 \times 10^{-2}$ \\
 & & 6.6\%
     & \cellcolor{green!20}$\mathbf{3.75 \times 10^{-3}}$ & \cellcolor{green!20}$\mathbf{5.66 \times 10^{-3}}$ & \cellcolor{green!20}$\mathbf{5.54 \times 10^{-3}}$ & \cellcolor{green!20}$\mathbf{1.03 \times 10^{-2}}$ & \cellcolor{green!20}$\mathbf{8.49 \times 10^{-3}}$ & \cellcolor{green!20}$\mathbf{1.04 \times 10^{-2}}$ & \cellcolor{green!20}$\mathbf{8.99 \times 10^{-3}}$ & \cellcolor{green!20}$\mathbf{6.09 \times 10^{-3}}$ & \cellcolor{green!20}$\mathbf{1.86 \times 10^{-2}}$ & \cellcolor{green!20}$\mathbf{4.52 \times 10^{-3}}$ \\
\bottomrule
\end{tabular}}}
\end{table*}

\begin{table}[t]
\centering
\small
\caption{Detection confidence ($\pval$) for $\Kirchenbauer$ and $\fedwm$ under $\randeigen$ as the global watermark ratio varies on $\alpaca$. Detection is aggregated over all client keys in this setting. $\Kirchenbauer$ remains undetectable, while $\fedwm$ generally possesses higher detection confidence as the global watermark ratio increases. Highlighted cells indicate $\pval < 0.05$.}
\label{tab:diffR}
{\setlength{\tabcolsep}{5pt}
\renewcommand{\arraystretch}{0.95}
\begin{tabular}{cc|cc}
\toprule
\textbf{Model} & \textbf{Global WM Ratio} & \textbf{$\Kirchenbauer$ }& \textbf{$\fedwm$} \\
\midrule
\multirow{3}{*}{$\pythia$}
& 6.6\%   & 0.417 & \cellcolor{green!20}$\mathbf{7.70 \times 10^{-4}}$ \\
& 10\%    & 0.329 & 0.104 \\
& 13.3\%  & 0.862 & \cellcolor{green!20}$\mathbf{2.51 \times 10^{-2}}$ \\
\midrule
\multirow{3}{*}{$\qwen$}
& 6.6\%   & 0.364                 & \cellcolor{green!20}$\mathbf{4.86 \times 10^{-7}}$ \\
& 10\%    & 0.535                 & \cellcolor{green!20}$\mathbf{1.32 \times 10^{-7}}$ \\
& 13.3\%  & $9.61 \times 10^{-2}$ & \cellcolor{green!20}$\mathbf{5.11 \times 10^{-11}}$ \\
\midrule
\multirow{3}{*}{$\llama$}
& 6.6\%   & 0.911 & \cellcolor{green!20}$\mathbf{3.28 \times 10^{-5}}$ \\
& 10\%    & 0.742 & \cellcolor{green!20}$\mathbf{2.38 \times 10^{-7}}$ \\
& 13.3\%  & 0.988 & \cellcolor{green!20}$\mathbf{8.37 \times 10^{-7}}$ \\
\bottomrule
\end{tabular}}
\end{table}

\begin{embox}
\textbf{(RQ2)} $\fedwm$ enables reliable black-box data provenance verification in federated learning under strong robust aggregators.
\end{embox}

\subsection{Robustness against Post-Training Attacks} 
\label{sec:eval-posttraining}
We evaluate two settings both using $\wmworker= 2$ watermarking clients, corresponding to $\wmratio = 6.6\%$. They differ by effective watermark ratio being 3.3\% or 6.6\%.

\myparatight{Effective Watermark Ratio of 3.3\%} Table~\ref{tab:post_attacks} demonstrates that at the most challenging effective watermark ratio of $3.3\%$, \qwen fine-tuned on watermarked \alpaca and \openweb remains robust across all evaluated attacks. \pythia is the most sensitive to post-training removal: it fails detection under second fine-tuning once training is sufficiently long ($\pval \geq 0.05$ on the same dataset beyond 5 epochs, or on a different dataset \oasst beyond 1 epoch), and under both evaluated pruning ratios on at least one dataset. \llama remains robust on \alpaca across all attacks, but loses detectability on \openweb under extended second fine-tuning (Diff~3e and beyond) and aggressive pruning ($r=40\%$). Pruning at $r=40\%$ is the most damaging attack overall, with the largest $\pval$ reaching 0.131. Distillation is the most robust attack: 5 of 6 model-dataset configurations remain detectable, with the only exception being \llama trained on \openweb ($\pval = 5.47 \times 10^{-2}$, just above the significance threshold).

\myparatight{Effective Watermark Ratio of 6.6\%} At a higher effective watermark fraction of 6.6\%, robustness improves substantially: all model-dataset-attack combinations remaining detectable (Table~\ref{tab:post_attacks}). In particular, \fedwm is robust to all fine-tuning settings. For instance, \qwen fine-tuned on \alpaca retains $\pval = 1.52 \times 10^{-3}$ even after the most aggressive fine-tuning attack (Diff 5e), and \llama fine-tuned on \alpaca yields $\pval = 3.93 \times 10^{-4}$ under the same condition. \fedwm is likewise robust to pruning, with 
$\pval$ reaching at most $1.86 \times 10^{-2}$ for \llama fine-tuned on \openweb even under the more aggressive pruning ratio of r=40\%. Finally, \fedwm is robust to all distillation attacks as well. For instance, \qwen on \alpaca retains $\pval = 7.97 \times 10^{-7}$ even after distillation.

\myparatight{Per LLM, per Dataset Observations} Overall, \pythia is the least robust model against watermark removal, likely due to its smaller capacity. With fewer parameters, it is more susceptible to forgetting the watermark signal when new knowledge is introduced. Robustness also varies by dataset: \alpaca , which is instruction-answer pairs dataset, is more resistant to post-training attacks than raw-text dataset, \openweb. This may be because instruction-answer pairs are more distinctively learned and thus harder to displace. Consistent with this, the $p$-values after initial training are already lower for \alpaca than for \openweb. More broadly, robustness against post-training attack depends not only on the attack type and model capacity, but also on the strength of the watermark signal at the end of the original training process.

\begin{embox}
\textbf{(RQ3)} $\fedwm$ is robust against most of the evaluated post-training watermark removal attacks.
\end{embox}

\subsection{Utility}
\begin{table*}[t]
\caption{Model utility after fine-tuning with natural data under $\fedavg$, $\Kirchenbauer$ watermarked data under $\fedavg$, $\Kirchenbauer$ watermarked data under $\randeigen$, and $\fedwm$ under $\randeigen$. Across entropy loss, C4~\cite{raffel2020exploring} perplexity, and MMLU-LLaMA~\cite{wang2024mmlu} accuracy, watermarking introduces only minor utility changes relative to the baseline of natural-data fine-tuning, indicating that both $\Kirchenbauer$ and \fedwm impose negligible cost on model quality.}
\label{tab:utility}
\centering
\small
{\setlength{\tabcolsep}{3pt}
\renewcommand{\arraystretch}{0.80}
\resizebox{\textwidth}{!}{
\begin{tabular}{cc|ccc|ccc|ccc|ccc}
\toprule
\textbf{Data} & \textbf{Model}
& \multicolumn{3}{c|}{\textbf{Natural data ($\fedavg$)}}
& \multicolumn{3}{c}{\textbf{$\Kirchenbauer$ data ($\fedavg$)}}
& \multicolumn{3}{c}{\textbf{$\Kirchenbauer$ data ($\randeigen$)}}
& \multicolumn{3}{c}{\textbf{$\fedwm$ ($\randeigen$)}} \\
\cmidrule(lr){3-5} \cmidrule(lr){6-8} \cmidrule(lr){9-11} \cmidrule(lr){12-14}
&
& \textbf{Loss $\downarrow$} & \makecell{\textbf{C4}\\PPL $\downarrow$} & \makecell{\textbf{MMLU}\\Acc $\uparrow$}
& \textbf{Loss $\downarrow$} & \makecell{\textbf{C4}\\PPL $\downarrow$} & \makecell{\textbf{MMLU}\\Acc $\uparrow$}
& \textbf{Loss $\downarrow$} & \makecell{\textbf{C4}\\PPL $\downarrow$} & \makecell{\textbf{MMLU}\\Acc $\uparrow$}
& \textbf{Loss $\downarrow$} & \makecell{\textbf{C4}\\PPL $\downarrow$} & \makecell{\textbf{MMLU}\\Acc $\uparrow$} \\
\midrule
$\alpaca$
& $\pythia$  & 1.246 & 2.032 & 0.267 & 1.246 & 2.012 & 0.264 & 1.251 & 2.027 & 0.271 & 1.254 & 2.031 & 0.268 \\
& $\qwen$    & 1.320 & 2.118 & 0.349 & 1.288 & 2.163 & 0.381 & 1.354 & 2.158 & 0.332 & 1.415 & 2.152 & 0.279 \\
& $\llama$   & 1.116 & 2.106 & 0.281 & 1.114 & 2.022 & 0.273 & 1.127 & 2.060 & 0.271 & 1.131 & 2.064 & 0.271 \\
\midrule
$\openweb$
& $\pythia$  & 2.739 & 1.942 & 0.262 & 2.740 & 1.941 & 0.263 & 2.743 & 1.943 & 0.263 & 2.744 & 1.943 & 0.261 \\
& $\qwen$    & 2.928 & 1.925 & 0.399 & 2.925 & 1.916 & 0.409 & 2.939 & 1.925 & 0.404 & 2.944 & 1.925 & 0.401 \\
& $\llama$   & 2.548 & 1.786 & 0.271 & 2.545 & 1.765 & 0.273 & 2.556 & 1.788 & 0.281 & 2.554 & 1.783 & 0.286 \\
\bottomrule
\end{tabular}}}
\end{table*}

In this section, we demonstrate that applying LLM watermarking, either $\Kirchenbauer$ or \fedwm, to fine-tuning data does not meaningfully hurt the utility of models trained on it. Specifically, Table~\ref{tab:utility} compares model performance when fine-tuning with 6.6\% of natural, $\Kirchenbauer$, and \fedwm data across multiple FL configurations.

\myparatight{KGW+ Has No Impact on Utility} Table~\ref{tab:utility} shows that $\Kirchenbauer$ has minimal impact on model utility. Across all three evaluation metrics, differences between the natural data baseline and the $\Kirchenbauer$ variants under \fedavg are very small. In some settings, $\Kirchenbauer$ even outperforms the natural data baseline. For instance, when fine-tuning \qwen on \openweb, $\Kirchenbauer$ data achieves better results on all three metrics comparing to natural data.
$\Kirchenbauer$ under \randeigen consistently shows slightly higher loss and perplexity than $\Kirchenbauer$ under \fedavg, indicating that \randeigen introduces a modest degradation in model utility (Table~\ref{tab:utility}). This is expected, as \randeigen does remove some client updates.

\myparatight{\fedwm Has No Negative Impact on Utility}
Overall, \fedwm's utility compared to baseline is similar. Although \fedwm under \randeigen generally produces higher loss and slightly worse metrics (Table~\ref{tab:utility}), the degradation remains minor. Compared to the baseline, C4 perplexity of \fedwm increases (worse utility) in roughly half the settings and decreases (better utility) in the other half, with a maximum increase of only $1.6\%$. We attribute this gap to \randeigen's over-filtering rather than to \fedwm. This is consistent with the original \randeigen paper, which reports less than $2\%$ accuracy degradation without poisoned samples.

\begin{embox}
\textbf{(RQ4)} Neither applying $\Kirchenbauer$ nor $\fedwm$ adversely impact model utility.
\end{embox}

\subsection{Ablation Study}

In this section, we vary multiple hyperparameters of the framework, demonstrating that $\fedwm$ remains effective across a wide range of hyperparameter settings. In section~\ref{sec:leave-one-out},  we further demonstrate the necessity of the proxy model: without it, $\fedwm$ becomes less effective in later rounds of FL fine-tuning.

\subsubsection{Global Watermark Ratio} \label{sec:eval-diffR}
\randeigen provides theoretical guarantee for $\wmratio \leq 8.33\%$, yet it also claims empirical effectiveness against attacks with higher $\wmratio$~\cite{lee2025practical}. We, therefore, evaluate $\Kirchenbauer$ and $\fedwm$ against \randeigen for global watermark ratio up to $13.3\%$. To evaluate higher global watermark ratio without re-generating the watermarked sets, we consider that clients share their keys and thus perform detection over a concatenated dataset. Under this mode, the effective watermark ratio is cumulated across clients and thus equals $\wmratio$. Table~\ref{tab:diffR} shows that detection of $\Kirchenbauer$ consistently fails across all evaluated $\wmratio$, with $\pval$ remaining close to $0.5$ after fine-tuning under \randeigen. Even at the highest ratio ($13.3\%$), where \qwen fine-tuned on \alpaca yields a slightly lower $\pval$, detection still fails to meet the predefined threshold $0.05$. Although $\wmratio$ exceeds the assumption underlying the theoretical guarantee, we do not observe a systematic decrease in $\pval$ as $\wmratio$ increases, indicating that \randeigen maintains strong empirical effectiveness within this range.

Therefore, the improved detectability of \fedwm at higher $\wmratio$ is not due to a weakening of \randeigen, but rather stems from stronger watermark signals. Specifically, a higher $\wmratio$ directly translates into a higher effective watermark ratio in this setting, allowing the watermark signal to accumulate across larger watermarked dataset and yield lower $\pval$. We evaluate our proposed scheme, \fedwm, under the same range of $\wmratio$. Table~\ref{tab:diffR} shows that \fedwm achieves reliable detection across almost all settings, with the best $\pval$ reaching $5.11 \times 10^{-11}$ for \qwen fine-tuned on \alpaca. Detectability generally improves with increasing effective watermark ratio for a given setting. For example, for \llama, $\pval$ decreases from $3.56 \times 10^{-3}$ to $8.37 \times 10^{-7}$ as effective watermark ratio increases from $3.33\%$ to $13.3\%$. The only exception is \pythia at $\wmratio = 10\%$, which yields a higher $\pval$ of $0.104$, likely due to model-specific variability and its smaller capacity for memorization, given its size.

\begin{figure*}[t]
\centering

\begin{subfigure}{0.325\textwidth}
    \centering
    \includegraphics[width=1.0\linewidth]{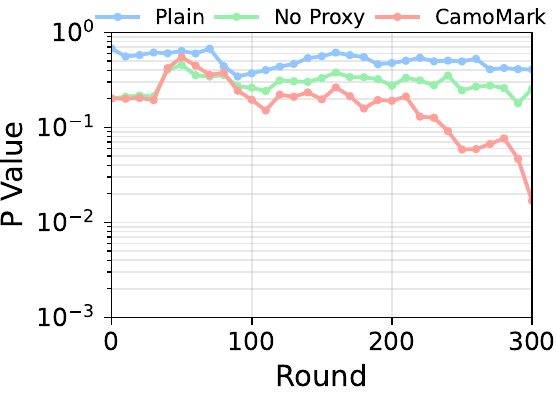}
    \caption{$\pval$ for $\pythia$}
\end{subfigure}
\begin{subfigure}{0.325\textwidth}
    \centering
    \includegraphics[width=1.0\linewidth]{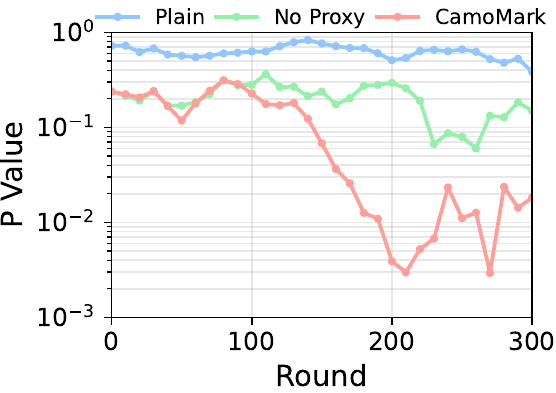}
    \caption{$\pval$ for $\qwen$}
\end{subfigure}
\begin{subfigure}{0.325\textwidth}
    \centering
    \includegraphics[width=1.0\linewidth]{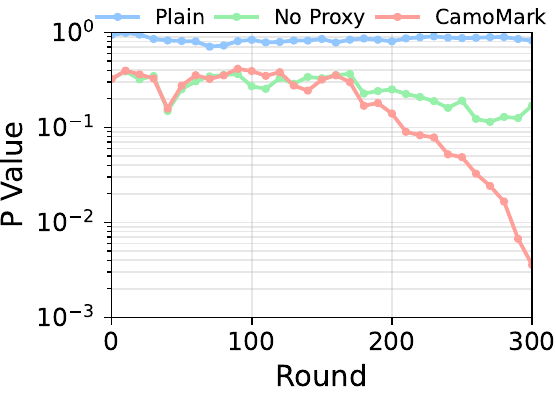}
    \caption{$\pval$ for $\llama$}
\end{subfigure}

\vspace{0.5em}

\begin{subfigure}{0.32\textwidth}
    \centering
    \includegraphics[width=1.0\linewidth]{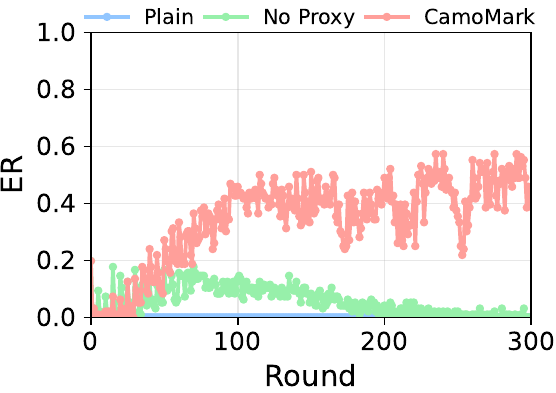}
    \caption{$\evasionrate$ for $\pythia$}
\end{subfigure}
\begin{subfigure}{0.32\textwidth}
    \centering
    \includegraphics[width=1.0\linewidth]{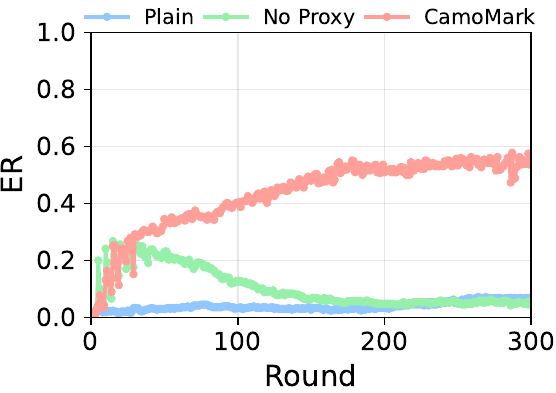}
    \caption{$\evasionrate$ for $\qwen$}
\end{subfigure}
\begin{subfigure}{0.32\textwidth}
    \centering
    \includegraphics[width=1.0\linewidth]{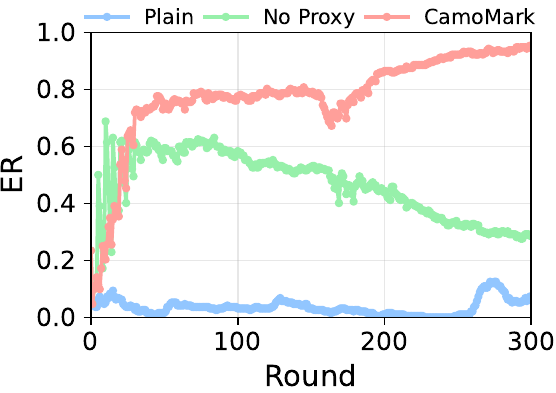}
    \caption{$\evasionrate$ for $\llama$}
\end{subfigure}

\caption{Detection confidence ($\pval$) and $\randeigen$ evasion rate ($\evasionrate$) over training in the leave-proxy-out experiment on $\alpaca$. $\Kirchenbauer$ is consistently filtered out, and $\fedwm$ achieves reliable detection only when the proxy model is used.}
\label{fig:six}
\end{figure*}

\subsubsection{Non-IID Client Data}
We evaluate \fedwm in a more realistic FL setting with non-independent and identically distributed (non-IID) client data. We use a state-of-the-art heterogeneous Byzantine ML defense~\cite{allouah2023fixing} that pre-processes updates via nearest-neighbor mixing (NNM)~\cite{allouah2023fixing}, then aggregates the mixed updates using Multi-Krum~\cite{blanchard2017machine}. To simulate data heterogeneity, following the same settings as~\cite{allouah2023fixing}, we treat \alpaca and \openweb as two distinct data sources and sample each client's mixture proportion from a Dirichlet distribution with concentration parameter $\alpha$~\cite{hsu2019measuring}. We consider three heterogeneity levels: extreme ($\alpha=0.1$), moderate ($\alpha=1$), and low ($\alpha=10$).

For $\pythia$ with the effective watermark ratio of $3.3\%$, $\Kirchenbauer$ again exhibits unreliable detection confidence, with \pvals of $0.544$, $0.282$, and $0.515$ across the three evaluated levels of data heterogeneity. In contrast, $\fedwm$ remains reliably detectable in all cases, with \pvals of $8.61 \times 10^{-5}$, $4.38 \times 10^{-3}$, and $1.00 \times 10^{-2}$ for $\alpha = 0.1$, $1$, and $10$, respectively, all falling below the detection threshold of $0.05$. This is expected, since weak robust aggregators such as Multi-Krum offer a much looser bias-bound guarantee than strong robust aggregators such as $\randeigen$. $\fedwm$ finds it considerably easier to reach the desired detection confidence in the non-IID setting with Multi-Krum than in the IID setting with $\randeigen$. Furthermore, detectability improves as $\alpha$ decreases, which corresponds to higher heterogeneity. This is a natural consequence of higher heterogeneity increasing the variance of the clean updates distribution, which allows the watermark to blend in more easily.

Strong robust aggregators such as $\randeigen$ remain underexplored for non-IID settings. Here, watermarked updates can potentially escape filtering because of the natural high variance of non-IID data, or the aggregator can over-filter useful updates. This is confirmed by our preliminary results, which show that \randeigen, under the same data heterogeneity settings as above, filters out all client updates in some communication rounds, preventing meaningful aggregation.

\subsubsection{Proxy model} 
\label{sec:leave-one-out}
We examine the importance of the proxy model to $\fedwm$ via a leave-one-out experiment. Figure~\ref{fig:six} reports the $\pval$ and $\evasionrate$ across communication rounds on $\alpaca$ for three configurations: (1) $\Kirchenbauer$, (2) complete $\fedwm$, and (3) $\fedwm$ without the proxy model, where the clean clients' gradient is instead estimated by running the global model on local natural data at every round.

The trend is consistent across all model configurations. $\Kirchenbauer$ alone yields persistently high $\pval$ and near-zero $\evasionrate$, confirming it is reliably suppressed by $\randeigen$. Complete $\fedwm$ becomes progressively more effective over time: $\evasionrate$ increases and $\pval$ decreases, indicating that as the global model accumulates watermark signal over time, it becomes harder for $\randeigen$ to suppress watermarked updates in later rounds. In contrast, $\fedwm$ without the proxy model shows an $\evasionrate$ pattern that initially rises until approximately round 50 before declining. Since evasion is never sustained, detection improves upon $\Kirchenbauer$ but consistently fails ($\pval > 0.05$) in the absence of the proxy model.

This degradation highlights the critical role of the proxy model, particularly in later rounds. As FL training progresses, the global model is continuously shaped by the clean clients' data distribution, but never by the watermarking client's clean data. Approximating the clean gradients from the global model therefore introduces a growing discrepancy: clean clients compute updates from a global model well-aligned with their data distribution, whereas the estimation is derived from watermark clients' clean data, which the global model has never been trained on. This causes the estimation to increasingly diverge from the true clean gradient distribution, causing proxy updates themselves to behave as outliers and undermine the optimization objective, especially for later rounds. 

Note that $\evasionrate$ does not reach $0.0\%$ even when detection fails and \pval remains high. For instance, \llama fine-tuned on \alpaca with no proxy model configuration demonstrates an $\evasionrate$ ranging from 20\% to 60\%. However, its \pval never decreases below $0.1$. This is because $\randeigen$ operates at the layer level, selectively removing only those layers that deviate significantly from clean updates, i.e., layers with strong watermark signals. Layers exhibiting weaker deviations, which carry less watermark-specific signal but more general knowledge, may survive filtering. Therefore, the global model can still learn general knowledge from watermarked clients through the fraction of surviving updates, as the residual watermark signal is too weak to enable post-training detection.

\begin{embox}
\textbf{(RQ5)}  A proxy model is essential for $\fedwm$ to function effectively. With it, $\fedwm$ achieves reliable black-box data provenance verification across diverse hyperparameter settings.
\end{embox}

\section{Related Work}
\myparatight{FL Training} 
Standard FL algorithms like FedSGD and FedAvg~\cite{mcmahan2017communication} converge slowly under non-IID data~\cite{karimireddy2020scaffold} or noisy environment~\cite{zhang2020adaptive}. To accelerate convergence, subsequent works employ adaptive local learning rate~\cite{xie2019local, sun2023fedlalr} and adaptive local interval~\cite{spiridonoff2020local, ji2020history}. Our FL setup adapts the FedOpt framework~\cite{reddi2020adaptive}, which utilizes adaptive optimizers to accelerate the convergence speed. 

\myparatight{Robust Aggregators and Byzantine Attacks}
Robust aggregators commonly defend against poisoning attacks by bounding the distance between the aggregated update and the mean of the clean updates. Weak robust aggregators~\cite{blanchard2017machine, chen2020distributed, yin2018byzantine} compute dimension-wise centrality, suffering from a worst-case bias of $O(\sqrt{\epsilon d})$~\cite{lai2016agnostic, 3ddcca11-deed-367a-9bc9-a521093d68cc}. Polynomial-time strong robust aggregators~\cite{diakonikolas2017being,hopkins2020robust,kothari2017outlier} provide dimension-independent bias bounds in theory but are computationally impractical at LLM scale~\cite{choudhary2024attacking,lee2025practical}.

Baruch et al.~\cite{baruch2019little} show that small, persistent update perturbations are sufficient to plant backdoors in FL. Shejwalkar and Houmansadr~\cite{shejwalkar2021manipulating} formulate an optimization problem for the scaling coefficient that maximizes damage to the global model. However, both works evaluate only against weak robust aggregators, whose bias guarantees scale with dimensionality and are therefore considerably looser. \hidra~\cite{choudhary2024attacking} instead exploits the gap between the theoretical and practical guarantees of strong robust aggregators, making them unreliable at filtering corrupted updates in practice. Specifically, \hidra crafts poisoned updates to remain just within a strong robust aggregator's practical bias bound, thereby avoiding removal.

The recent work \randeigen~\cite{lee2025practical} is, to our knowledge, the first method to bridge this gap. It attains a dimension-independent bias bound in practice with quasi-linear runtime, and it is robust to \hidra. Because \hidra-style attacks cannot circumvent \randeigen's filtering, they cannot be directly repurposed as watermarking mechanisms either. Therefore, we evaluate \fedwm against it as the current representative of this emerging class of practically-strong aggregators.

\myparatight{Gradient Matching} Geiping et al.~\cite{geiping2020witches} use a similar idea of gradient matching to craft clean-label examples whose induced gradient aligns with a targeted mis-classification objective. However, their threat model is different from ours: centralized rather than federated, with no server aggregation step to evade. Moreover, \fedwm gradients are not intended to misclassify or reduce accuracy.

\myparatight{LLM Watermarking} LLM watermarking embed a hidden, statistically detectable signal into LLM-generated text without noticeably degrading text quality, enabling reliable distinction between AI-generated and human-written text~\cite{kirchenbauer2023watermark, aaronson2023watermarking, christ2024undetectable, kuditipudi2023robust, zhao2023provable}. Existing schemes differ mainly in how they introduce randomness during decoding: hashing-based methods (\Kirchenbauer~\cite{kirchenbauer2023watermark}, \Aarson~\cite{aaronson2023watermarking}) derive randomness from the local token context, while key-sequence methods (\Kuditipudi~\cite{kuditipudi2023robust}) fix randomness via a predefined sequence. This distinction leads to different trade-offs between distortion-freeness and robustness to repeated context (see Appendix~\ref{append:llm-watermarking} for details).
Subsequent works~\cite{sander2024watermarking, gu2023learnability} demonstrate that many LLM watermarks exhibit radioactivity in centralized setting. That is, if a model has fine-tuned with watermarked dataset, the watermark signal would persist in the output of such model, and can therefore be detected. What's more, such watermark is robust towards some watermark removal attack, such as further fine-tuning on clean dataset. We refer to this property as \textit{detectability} throughout the paper, which enables LLM watermarks to be used for model ownership verification.

\myparatight{Model Provenance}
Model provenance aims to determine whether a suspect model is derived from, copied from, or trained using a protected source. Early neural-network provenance methods often rely on model watermarking: Uchida et al. embed watermarks into model parameters via regularization for white-box ownership verification~\cite{uchida2017embedding}, while Zhang et al. and Adi et al. use secret trigger inputs with pre-specified outputs for black-box, backdoor-based verification~\cite{zhang2018protecting,adi2018turning}. These methods are influential for model-IP protection, but mainly target centralized training and classification models. Moreover, backdoor-based provenance introduces anomalous trigger-response behavior, which is conceptually similar to the outlier signals that robust aggregation is designed to remove.

Recent work extends provenance and fingerprinting to LLMs. ProFLingo generates black-box queries that elicit model-specific responses and tests whether a suspect model preserves them~\cite{jin2024proflingo}. LLMmap fingerprints deployed LLM applications via query-response behavior to infer the underlying model and version~\cite{pasquini2025llmmap}, while model provenance testing formulates model derivation as a statistical hypothesis test over black-box outputs~\cite{nikolic2025model}. These passive methods are attractive since they leave the model and training data untouched. However, they reveal only model identity or lineage, not whether a specific client’s data contributed to federated fine-tuning.

Backdoor-based provenance is closer to our setting because it actively embeds a verifiable signal during training. In FL, WAFFLE and FedGMark similarly use client-side backdoor watermarks for ownership verification~\cite{yang2023watermarking,yang2024fedgmark}. However, these methods target image or graph classification and rely on trigger-label behavior. Our work instead studies LLM fine-tuning data provenance: a client verifies via black-box access whether the final LLM was trained on its watermarked text. Unlike passive fingerprinting, which identifies model lineage, or backdoor watermarking, which embeds explicit trigger behavior, \fedwm uses LLM text watermarks as data-level provenance signals and optimizes the induced client updates to remain inliers under strong Byzantine-robust aggregation.

\section{Conclusion}

We present \fedwm, an aggregator-agnostic watermarking method for verifying model ownership in federated LLM fine-tuning under strong Byzantine-robust aggregation. Our results show that existing LLM watermarks remain detectable under standard averaging but are effectively removed by strong robust aggregator \randeigen. \fedwm addresses this limitation by optimizing watermarked updates to behave as inliers while preserving sufficient watermark signal for black-box detection. Across multiple models and datasets, \fedwm achieves reliable detection, remains robust to evaluated post-training removal attacks, and introduces only minor utility degradation. These findings highlight not only a previously overlooked vulnerability of watermarking in federated fine-tuning but also a practical path toward robust provenance verification for data-contributing clients.

\section*{Ethics Considerations}
This work studies data watermarking for federated fine-tuning of large language models (LLMs), with the goal of helping data-contributing clients verify whether a suspect model was trained on their watermarked data. Because the work concerns data provenance, model ownership, watermarking, and adversarial federated learning, we considered its ethical implications through a stakeholder-based analysis guided by the Menlo Report principles of Beneficence, Respect for Persons, Justice, and Respect for Law and Public Interest.

\myparatight{Stakeholders and benefits}
The primary stakeholders are clients who contribute data to federated learning. These clients may consent to training under specific conditions but later lack visibility into how the resulting model is used. \fedwm is intended to support such clients by enabling black-box verification of whether a model was trained on their watermarked data, even when a malicious server attempts to remove the watermark through Byzantine-robust aggregation. This can improve accountability, support data provenance, and help data owners detect potential violations of agreed data-use conditions. Other stakeholders include clean clients, federated-learning servers, model developers, end users, dataset creators, and the broader research community. These groups may benefit from stronger provenance mechanisms, but they may also be affected if watermarking degrades model utility or if detection evidence is misinterpreted.

\myparatight{Potential harms}
The main ethical risks are false attribution, failed detection, and dual use. A false positive could incorrectly suggest that a model was trained on a client's data, while a false negative could fail to detect misuse. In addition, showing that strong robust aggregators can remove existing watermark signals may inform malicious servers seeking to evade provenance checks. Watermarking could also be misused to make unsupported ownership claims over data or models. Finally, if watermarking degraded the trained model, downstream users could be affected by reduced model quality.

\myparatight{Mitigations}
We mitigate these risks in several ways. First, we report watermark detection as statistical evidence using $p$-values rather than as definitive proof of ownership or legal violation. Second, we evaluate \fedwm under both in-training removal attacks and post-training attacks, including continued fine-tuning, pruning, and distillation. Third, we evaluate model utility and show that \fedwm preserves performance in the studied settings. Fourth, our method does not require access to other clients' private data; watermarking clients use their own data and secret keys, and clean-gradient estimation is performed locally. We therefore frame \fedwm as one component of a broader provenance and governance process, not as a standalone authority for ownership disputes.

\myparatight{Decision}
We believe the benefits of this research outweigh the risks. Federated LLM training creates an accountability gap: clients may contribute data while having limited ability to verify downstream misuse of the resulting model. Existing LLM watermarking methods can fail under strong robust aggregation, which means data contributors may have a false sense of protection. Publishing this work helps the community understand that limitation and develop stronger provenance mechanisms. Although the work has dual-use aspects, the adversarial capability we study---using robust aggregation to filter anomalous updates---is already available in practical federated-learning systems. We therefore judge that disclosing the weakness and proposing a defense better serves the public interest than leaving the issue unexamined.

\bibliographystyle{IEEEtran}
\bibliography{references}

@String{Computing = "Computing" }

@String{Computer = "{IEEE} Computer" }

@String{Springer = "Springer-Verlag" }

@article{luo2023wizardcoder,
  title={Wizardcoder: Empowering code large language models with evol-instruct},
  author={Luo, Ziyang and Xu, Can and Zhao, Pu and Sun, Qingfeng and Geng, Xiubo and Hu, Wenxiang and Tao, Chongyang and Ma, Jing and Lin, Qingwei and Jiang, Daxin},
  journal={arXiv preprint arXiv:2306.08568},
  year={2023}
}

@article{assaad2024meta,
  author  = {Zena Assaad},
  title   = {Meta now allows military agencies to access its AI software. It poses a moral dilemma for everybody who uses it},
  journal = {The Conversation},
  year    = {2024},
  month   = {November},
  url     = {https://theconversation.com/meta-now-allows-military-agencies-to-access-its-ai-software-it-poses-a-moral-dilemma-for-everybody-who-uses-it-243250}
}

@misc{longpre2023dataprovenanceinitiativelarge,
      title={The Data Provenance Initiative: A Large Scale Audit of Dataset Licensing \& Attribution in AI}, 
      author={Shayne Longpre and Robert Mahari and Anthony Chen and Naana Obeng-Marnu and Damien Sileo and William Brannon and Niklas Muennighoff and Nathan Khazam and Jad Kabbara and Kartik Perisetla and Xinyi Wu and Enrico Shippole and Kurt Bollacker and Tongshuang Wu and Luis Villa and Sandy Pentland and Sara Hooker},
      year={2023},
      eprint={2310.16787},
      archivePrefix={arXiv},
      primaryClass={cs.CL},
      url={https://arxiv.org/abs/2310.16787}, 
}

@article{yang2024fedgmark,
  title={Fedgmark: Certifiably robust watermarking for federated graph learning},
  author={Yang, Yuxin and Li, Qiang and Hong, Yuan and Wang, Binghui},
  journal={Advances in Neural Information Processing Systems},
  volume={37},
  pages={48971--48995},
  year={2024}
}

@inproceedings{kirchenbauer2023watermark,
  title={A watermark for large language models},
  author={Kirchenbauer, John and Geiping, Jonas and Wen, Yuxin and Katz, Jonathan and Miers, Ian and Goldstein, Tom},
  booktitle={International conference on machine learning},
  pages={17061--17084},
  year={2023},
  organization={PMLR}
}

@misc{aaronson2023watermarking,
  author       = {Scott Aaronson and Hendrik Kirchner},
  title        = {Watermarking {GPT} Outputs},
  year         = {2023},
  howpublished = {\url{https://scottaaronson.blog/?m=202302}},
}

@article{kuditipudi2023robust,
  title={Robust distortion-free watermarks for language models},
  author={Kuditipudi, Rohith and Thickstun, John and Hashimoto, Tatsunori and Liang, Percy},
  journal={arXiv preprint arXiv:2307.15593},
  year={2023}
}

@inproceedings{christ2024undetectable,
  title={Undetectable watermarks for language models},
  author={Christ, Miranda and Gunn, Sam and Zamir, Or},
  booktitle={The Thirty Seventh Annual Conference on Learning Theory},
  pages={1125--1139},
  year={2024},
  organization={PMLR}
}

@article{zhao2023provable,
  title={Provable robust watermarking for ai-generated text},
  author={Zhao, Xuandong and Ananth, Prabhanjan and Li, Lei and Wang, Yu-Xiang},
  journal={arXiv preprint arXiv:2306.17439},
  year={2023}
}

@article{hu2024stable,
  title={Stable signature is unstable: Removing image watermark from diffusion models},
  author={Hu, Yuepeng and Jiang, Zhengyuan and Guo, Moyang and Gong, Neil},
  journal={arXiv preprint arXiv:2405.07145},
  year={2024}
}

@article{sander2024watermarking,
  title={Watermarking makes language models radioactive},
  author={Sander, Tom and Fernandez, Pierre and Durmus, Alain and Douze, Matthijs and Furon, Teddy},
  journal={Advances in Neural Information Processing Systems},
  volume={37},
  pages={21079--21113},
  year={2024}
}

@article{gu2023learnability,
  title={On the learnability of watermarks for language models},
  author={Gu, Chenchen and Li, Xiang Lisa and Liang, Percy and Hashimoto, Tatsunori},
  journal={arXiv preprint arXiv:2312.04469},
  year={2023}
}

@inproceedings{chen2022copy,
  title={Copy, right? a testing framework for copyright protection of deep learning models},
  author={Chen, Jialuo and Wang, Jingyi and Peng, Tinglan and Sun, Youcheng and Cheng, Peng and Ji, Shouling and Ma, Xingjun and Li, Bo and Song, Dawn},
  booktitle={2022 IEEE symposium on security and privacy (SP)},
  pages={824--841},
  year={2022},
  organization={IEEE}
}

@inproceedings{karimireddy2020scaffold,
  title={Scaffold: Stochastic controlled averaging for federated learning},
  author={Karimireddy, Sai Praneeth and Kale, Satyen and Mohri, Mehryar and Reddi, Sashank and Stich, Sebastian and Suresh, Ananda Theertha},
  booktitle={International conference on machine learning},
  pages={5132--5143},
  year={2020},
  organization={PMLR}
}

@article{xie2019local,
  title={Local adaalter: Communication-efficient stochastic gradient descent with adaptive learning rates},
  author={Xie, Cong and Koyejo, Oluwasanmi and Gupta, Indranil and Lin, Haibin},
  journal={arXiv preprint arXiv:1911.09030},
  year={2019}
}

@article{yang2023watermarking,
  title={Watermarking in secure federated learning: A verification framework based on client-side backdooring},
  author={Yang, Wenyuan and Shao, Shuo and Yang, Yue and Liu, Xiyao and Liu, Ximeng and Xia, Zhihua and Schaefer, Gerald and Fang, Hui},
  journal={ACM Transactions on Intelligent Systems and Technology},
  volume={15},
  number={1},
  pages={1--25},
  year={2023},
  publisher={ACM New York, NY}
}

@article{zhang2020adaptive,
  title={Why are adaptive methods good for attention models?},
  author={Zhang, Jingzhao and Karimireddy, Sai Praneeth and Veit, Andreas and Kim, Seungyeon and Reddi, Sashank and Kumar, Sanjiv and Sra, Suvrit},
  journal={Advances in Neural Information Processing Systems},
  volume={33},
  pages={15383--15393},
  year={2020}
}

@article{sun2023fedlalr,
  title={Fedlalr: Client-specific adaptive learning rates achieve linear speedup for non-iid data},
  author={Sun, Hao and Shen, Li and Chen, Shixiang and Sun, Jingwei and Li, Jing and Sun, Guangzhong and Tao, Dacheng},
  journal={arXiv preprint arXiv:2309.09719},
  year={2023}
}

@article{spiridonoff2020local,
  title={Local sgd with a communication overhead depending only on the number of workers},
  author={Spiridonoff, Artin and Olshevsky, Alex and Paschalidis, Ioannis Ch},
  journal={arXiv preprint arXiv:2006.02582},
  year={2020}
}

@inproceedings{ji2020history,
  title={History-gradient aided batch size adaptation for variance reduced algorithms},
  author={Ji, Kaiyi and Wang, Zhe and Weng, Bowen and Zhou, Yi and Zhang, Wei and Liang, Yingbin},
  booktitle={International Conference on Machine Learning},
  pages={4762--4772},
  year={2020},
  organization={PMLR}
}

@article{reddi2020adaptive,
  title={Adaptive federated optimization},
  author={Reddi, Sashank and Charles, Zachary and Zaheer, Manzil and Garrett, Zachary and Rush, Keith and Kone{\v{c}}n{\`y}, Jakub and Kumar, Sanjiv and McMahan, H Brendan},
  journal={arXiv preprint arXiv:2003.00295},
  year={2020}
}

@inproceedings{lai2016agnostic,
  title={Agnostic estimation of mean and covariance},
  author={Lai, Kevin A and Rao, Anup B and Vempala, Santosh},
  booktitle={2016 IEEE 57th Annual Symposium on Foundations of Computer Science (FOCS)},
  pages={665--674},
  year={2016},
  organization={IEEE}
}

@article{3ddcca11-deed-367a-9bc9-a521093d68cc,
     ISSN = {00905364, 21688966},
     URL = {https://www.jstor.org/stable/27028779},
     author = {Gábor Lugosi and Shahar Mendelson},
     journal = {The Annals of Statistics},
     number = {1},
     pages = {pp. 393--410},
     publisher = {Institute of Mathematical Statistics},
     title = {ROBUST MULTIVARIATE MEAN ESTIMATION: THE OPTIMALITY OF TRIMMED MEAN},
     urldate = {2025-07-30},
     volume = {49},
     year = {2021}
}

@article{chen2020distributed,
  title={Distributed training with heterogeneous data: Bridging median-and mean-based algorithms},
  author={Chen, Xiangyi and Chen, Tiancong and Sun, Haoran and Wu, Steven Z and Hong, Mingyi},
  journal={Advances in Neural Information Processing Systems},
  volume={33},
  pages={21616--21626},
  year={2020}
}

@article{diakonikolas2019recent,
  title={Recent advances in algorithmic high-dimensional robust statistics},
  author={Diakonikolas, Ilias and Kane, Daniel M},
  journal={arXiv preprint arXiv:1911.05911},
  year={2019}
}

@inproceedings{diakonikolas2017being,
  title={Being robust (in high dimensions) can be practical},
  author={Diakonikolas, Ilias and Kamath, Gautam and Kane, Daniel M and Li, Jerry and Moitra, Ankur and Stewart, Alistair},
  booktitle={International Conference on Machine Learning},
  pages={999--1008},
  year={2017},
  organization={PMLR}
}

@article{hopkins2020robust,
  title={Robust and heavy-tailed mean estimation made simple, via regret minimization},
  author={Hopkins, Sam and Li, Jerry and Zhang, Fred},
  journal={Advances in Neural Information Processing Systems},
  volume={33},
  pages={11902--11912},
  year={2020}
}

@article{kothari2017outlier,
  title={Outlier-robust moment-estimation via sum-of-squares},
  author={Kothari, Pravesh K and Steurer, David},
  journal={arXiv preprint arXiv:1711.11581},
  year={2017}
}

@incollection{tukey1960contaminated,
  author    = {John W. Tukey},
  title     = {A Survey of Sampling from Contaminated Distributions},
  booktitle = {Contributions to Probability and Statistics},
  publisher = {Stanford University Press},
  year      = {1960},
  pages     = {448--485}
}

@inproceedings{choudhary2024attacking,
  title={Attacking byzantine robust aggregation in high dimensions},
  author={Choudhary, Sarthak and Kolluri, Aashish and Saxena, Prateek},
  booktitle={2024 IEEE Symposium on Security and Privacy (SP)},
  pages={1325--1344},
  year={2024},
  organization={IEEE}
}

@inproceedings{lee2025practical,
  title={A Practical and Secure Byzantine Robust Aggregator},
  author={Lee, De Zhang and Kolluri, Aashish and Saxena, Prateek and Chang, Ee-Chien},
  booktitle={Proceedings of the 2025 ACM SIGSAC Conference on Computer and Communications Security},
  pages={2877--2891},
  year={2025}
}

@article{wu2025survey,
  title={A survey on federated fine-tuning of large language models},
  author={Wu, Yebo and Tian, Chunlin and Li, Jingguang and Sun, He and Tam, Kahou and Zhou, Zhanting and Liao, Haicheng and Xiong, Jing and Guo, Zhijiang and Li, Li and others},
  journal={arXiv preprint arXiv:2503.12016},
  year={2025}
}

@article{longpre2024data,
  title={Data authenticity, consent, and provenance for AI are all broken: What will it take to fix them?},
  author={Longpre, Shayne and Mahari, Robert and Obeng-Marnu, Naana and Brannon, William and South, Tobin and Kabbara, Jad and Pentland, Sandy},
  year={2024},
  publisher={MIT}
}

@article{shao2024fedtracker,
  title={Fedtracker: Furnishing ownership verification and traceability for federated learning model},
  author={Shao, Shuo and Yang, Wenyuan and Gu, Hanlin and Qin, Zhan and Fan, Lixin and Yang, Qiang and Ren, Kui},
  journal={IEEE Transactions on Dependable and Secure Computing},
  volume={22},
  number={1},
  pages={114--131},
  year={2024},
  publisher={IEEE}
}

@article{li2022fedipr,
  title={FedIPR: Ownership verification for federated deep neural network models},
  author={Li, Bowen and Fan, Lixin and Gu, Hanlin and Li, Jie and Yang, Qiang},
  journal={IEEE Transactions on Pattern Analysis and Machine Intelligence},
  volume={45},
  number={4},
  pages={4521--4536},
  year={2022},
  publisher={IEEE}
}

@inproceedings{adi2018turning,
  title={Turning your weakness into a strength: Watermarking deep neural networks by backdooring},
  author={Adi, Yossi and Baum, Carsten and Cisse, Moustapha and Pinkas, Benny and Keshet, Joseph},
  booktitle={27th USENIX security symposium (USENIX Security 18)},
  pages={1615--1631},
  year={2018}
}

@inproceedings{cui2025robust,
  title={Robust data watermarking in language models by injecting fictitious knowledge},
  author={Cui, Xinyue and Wei, Johnny and Swayamdipta, Swabha and Jia, Robin},
  booktitle={Findings of the Association for Computational Linguistics: ACL 2025},
  pages={14292--14306},
  year={2025}
}

@inproceedings{wei2024proving,
  title={Proving membership in LLM pretraining data via data watermarks},
  author={Wei, Johnny and Wang, Ryan and Jia, Robin},
  booktitle={Findings of the Association for Computational Linguistics: ACL 2024},
  pages={13306--13320},
  year={2024}
}

@inproceedings{jia2021entangled,
  title={Entangled watermarks as a defense against model extraction},
  author={Jia, Hengrui and Choquette-Choo, Christopher A and Chandrasekaran, Varun and Papernot, Nicolas},
  booktitle={30th USENIX security symposium (USENIX Security 21)},
  pages={1937--1954},
  year={2021}
}

@article{wang2024mmlu,
  title={Mmlu-pro: A more robust and challenging multi-task language understanding benchmark},
  author={Wang, Yubo and Ma, Xueguang and Zhang, Ge and Ni, Yuansheng and Chandra, Abhranil and Guo, Shiguang and Ren, Weiming and Arulraj, Aaran and He, Xuan and Jiang, Ziyan and others},
  journal={Advances in Neural Information Processing Systems},
  volume={37},
  pages={95266--95290},
  year={2024}
}

@article{raffel2020exploring,
  title={Exploring the limits of transfer learning with a unified text-to-text transformer},
  author={Raffel, Colin and Shazeer, Noam and Roberts, Adam and Lee, Katherine and Narang, Sharan and Matena, Michael and Zhou, Yanqi and Li, Wei and Liu, Peter J},
  journal={Journal of machine learning research},
  volume={21},
  number={140},
  pages={1--67},
  year={2020}
}

@inproceedings{uchida2017embedding,
  title={Embedding watermarks into deep neural networks},
  author={Uchida, Yusuke and Nagai, Yuki and Sakazawa, Shigeyuki and Satoh, Shin'ichi},
  booktitle={Proceedings of the 2017 ACM on international conference on multimedia retrieval},
  pages={269--277},
  year={2017}
}

@inproceedings{jin2024proflingo,
  title={Proflingo: A fingerprinting-based intellectual property protection scheme for large language models},
  author={Jin, Heng and Zhang, Chaoyu and Shi, Shanghao and Lou, Wenjing and Hou, Y Thomas},
  booktitle={2024 IEEE Conference on Communications and Network Security (CNS)},
  pages={1--9},
  year={2024},
  organization={IEEE}
}

@inproceedings{pasquini2025llmmap,
  title={$\{$LLMmap$\}$: Fingerprinting for large language models},
  author={Pasquini, Dario and Kornaropoulos, Evgenios M and Ateniese, Giuseppe},
  booktitle={34th USENIX Security Symposium (USENIX Security 25)},
  pages={299--318},
  year={2025}
}

@inproceedings{nikolic2025model,
  title={Model Provenance Testing for Large Language Models},
  author={Nikolic, Ivica and Baluta, Teodora and Saxena, Prateek},
  booktitle={Advances in Neural Information Processing Systems},
  year={2025}
}

@inproceedings{zhang2018protecting,
  title={Protecting intellectual property of deep neural networks with watermarking},
  author={Zhang, Jialong and Gu, Zhongshu and Jang, Jiyong and Wu, Hui and Stoecklin, Marc Ph and Huang, Heqing and Molloy, Ian},
  booktitle={Proceedings of the 2018 on Asia conference on computer and communications security},
  pages={159--172},
  year={2018}
}

@inproceedings{biderman2023pythia,
  title={Pythia: A suite for analyzing large language models across training and scaling},
  author={Biderman, Stella and Schoelkopf, Hailey and Anthony, Quentin Gregory and Bradley, Herbie and O’Brien, Kyle and Hallahan, Eric and Khan, Mohammad Aflah and Purohit, Shivanshu and Prashanth, USVSN Sai and Raff, Edward and others},
  booktitle={International conference on machine learning},
  pages={2397--2430},
  year={2023},
  organization={PMLR}
}

@article{touvron2023llama,
  title={Llama: Open and efficient foundation language models},
  author={Touvron, Hugo and Lavril, Thibaut and Izacard, Gautier and Martinet, Xavier and Lachaux, Marie-Anne and Lacroix, Timoth{\'e}e and Rozi{\`e}re, Baptiste and Goyal, Naman and Hambro, Eric and Azhar, Faisal and others},
  journal={arXiv preprint arXiv:2302.13971},
  year={2023}
}

@article{bai2023qwen,
  title={Qwen technical report},
  author={Bai, Jinze and Bai, Shuai and Chu, Yunfei and Cui, Zeyu and Dang, Kai and Deng, Xiaodong and Fan, Yang and Ge, Wenbin and Han, Yu and Huang, Fei and others},
  journal={arXiv preprint arXiv:2309.16609},
  year={2023}
}

@misc{taori2023stanford,
  title={Stanford alpaca: An instruction-following llama model},
  author={Taori, Rohan and Gulrajani, Ishaan and Zhang, Tianyi and Dubois, Yann and Li, Xuechen and Guestrin, Carlos and Liang, Percy and Hashimoto, Tatsunori B},
  year={2023},
  publisher={Stanford, CA, USA}
}

@misc{Gokaslan2019OpenWeb,  
	title={OpenWebText Corpus},
	author={Aaron Gokaslan and Vanya Cohen},
	howpublished={\url{http://Skylion007.github.io/OpenWebTextCorpus}}, 
	year={2019}
}

@article{ouyang2022training,
  title={Training language models to follow instructions with human feedback},
  author={Ouyang, Long and Wu, Jeffrey and Jiang, Xu and Almeida, Diogo and Wainwright, Carroll and Mishkin, Pamela and Zhang, Chong and Agarwal, Sandhini and Slama, Katarina and Ray, Alex and others},
  journal={Advances in neural information processing systems},
  volume={35},
  pages={27730--27744},
  year={2022}
}

@article{wei2021finetuned,
  title={Finetuned language models are zero-shot learners},
  author={Wei, Jason and Bosma, Maarten and Zhao, Vincent Y and Guu, Kelvin and Yu, Adams Wei and Lester, Brian and Du, Nan and Dai, Andrew M and Le, Quoc V},
  journal={arXiv preprint arXiv:2109.01652},
  year={2021}
}

@article{li2021survey,
  title={A survey on federated learning systems: Vision, hype and reality for data privacy and protection},
  author={Li, Qinbin and Wen, Zeyi and Wu, Zhaomin and Hu, Sixu and Wang, Naibo and Li, Yuan and Liu, Xu and He, Bingsheng},
  journal={IEEE Transactions on Knowledge and Data Engineering},
  volume={35},
  number={4},
  pages={3347--3366},
  year={2021},
  publisher={IEEE}
}

@article{rieke2020future,
  title={The future of digital health with federated learning},
  author={Rieke, Nicola and Hancox, Jonny and Li, Wenqi and Milletari, Fausto and Roth, Holger R and Albarqouni, Shadi and Bakas, Spyridon and Galtier, Mathieu N and Landman, Bennett A and Maier-Hein, Klaus and others},
  journal={NPJ digital medicine},
  volume={3},
  number={1},
  pages={119},
  year={2020},
  publisher={Nature Publishing Group UK London}
}

@inproceedings{zhang2024towards,
  title={Towards building the federatedgpt: Federated instruction tuning},
  author={Zhang, Jianyi and Vahidian, Saeed and Kuo, Martin and Li, Chunyuan and Zhang, Ruiyi and Yu, Tong and Wang, Guoyin and Chen, Yiran},
  booktitle={ICASSP 2024-2024 IEEE international conference on acoustics, speech and signal processing (ICASSP)},
  pages={6915--6919},
  year={2024},
  organization={IEEE}
}

@inproceedings{mcmahan2017communication,
  title={Communication-efficient learning of deep networks from decentralized data},
  author={McMahan, Brendan and Moore, Eider and Ramage, Daniel and Hampson, Seth and y Arcas, Blaise Aguera},
  booktitle={Artificial intelligence and statistics},
  pages={1273--1282},
  year={2017},
  organization={PMLR}
}

@article{li2023black,
  title={Black-box dataset ownership verification via backdoor watermarking},
  author={Li, Yiming and Zhu, Mingyan and Yang, Xue and Jiang, Yong and Wei, Tao and Xia, Shu-Tao},
  journal={IEEE Transactions on Information Forensics and Security},
  volume={18},
  pages={2318--2332},
  year={2023},
  publisher={IEEE}
}

@article{blanchard2017machine,
  title={Machine learning with adversaries: Byzantine tolerant gradient descent},
  author={Blanchard, Peva and El Mhamdi, El Mahdi and Guerraoui, Rachid and Stainer, Julien},
  journal={Advances in neural information processing systems},
  volume={30},
  year={2017}
}

@inproceedings{yin2018byzantine,
  title={Byzantine-robust distributed learning: Towards optimal statistical rates},
  author={Yin, Dong and Chen, Yudong and Kannan, Ramchandran and Bartlett, Peter},
  booktitle={International conference on machine learning},
  pages={5650--5659},
  year={2018},
  organization={PMLR}
}

@inproceedings{shafieinejad2021robustness,
  title={On the robustness of backdoor-based watermarking in deep neural networks},
  author={Shafieinejad, Masoumeh and Lukas, Nils and Wang, Jiaqi and Li, Xinda and Kerschbaum, Florian},
  booktitle={Proceedings of the 2021 ACM workshop on information hiding and multimedia security},
  pages={177--188},
  year={2021}
}

@article{liu2018rethinking,
  title={Rethinking the value of network pruning},
  author={Liu, Zhuang and Sun, Mingjie and Zhou, Tinghui and Huang, Gao and Darrell, Trevor},
  journal={arXiv preprint arXiv:1810.05270},
  year={2018}
}

@article{renda2020comparing,
  title={Comparing rewinding and fine-tuning in neural network pruning},
  author={Renda, Alex and Frankle, Jonathan and Carbin, Michael},
  journal={arXiv preprint arXiv:2003.02389},
  year={2020}
}

@misc{frankle2023tweet,
  author       = {Jonathan Frankle},
  title        = {{Tweet [MosaicML]}},
  howpublished = {Twitter},
  year         = {2023},
  month        = {May},
  note         = {\url{https://twitter.com/jefrankle/status/1654848529834078208}}
}

@article{gebru2021datasheets,
  title={Datasheets for datasets},
  author={Gebru, Timnit and Morgenstern, Jamie and Vecchione, Briana and Vaughan, Jennifer Wortman and Wallach, Hanna and Iii, Hal Daum{\'e} and Crawford, Kate},
  journal={Communications of the ACM},
  volume={64},
  number={12},
  pages={86--92},
  year={2021},
  publisher={ACM New York, NY, USA}
}

@article{longpre2024consent,
  title={Consent in crisis: The rapid decline of the ai data commons, 2024},
  author={Longpre, Shayne and Mahari, Robert and Lee, Ariel and Lund, Campbell and Oderinwale, Hamidah and Brannon, William and Saxena, Nayan and Obeng-Marnu, Naana and South, Tobin and Hunter, Cole and others},
  journal={URL https://arxiv. org/abs/2407.14933},
  year={2024}
}

@article{kairouz2021advances,
  title={Advances and open problems in federated learning},
  author={Kairouz, Peter and McMahan, H Brendan},
  journal={Foundations and trends in machine learning},
  volume={14},
  number={1-2},
  pages={1--210},
  year={2021},
  publisher={Emerald Publishing Limited}
}

@article{hu2022lora,
  title={Lora: Low-rank adaptation of large language models.},
  author={Hu, Edward J and Shen, Yelong and Wallis, Phillip and Allen-Zhu, Zeyuan and Li, Yuanzhi and Wang, Shean and Wang, Liang and Chen, Weizhu and others},
  journal={ICLR},
  volume={1},
  number={2},
  pages={3},
  year={2022}
}

@article{gao2020pile,
  title={The pile: An 800gb dataset of diverse text for language modeling},
  author={Gao, Leo and Biderman, Stella and Black, Sid and Golding, Laurence and Hoppe, Travis and Foster, Charles and Phang, Jason and He, Horace and Thite, Anish and Nabeshima, Noa and others},
  journal={arXiv preprint arXiv:2101.00027},
  year={2020}
}

@article{lyu2020towards,
  title={Towards fair and privacy-preserving federated deep models},
  author={Lyu, Lingjuan and Yu, Jiangshan and Nandakumar, Karthik and Li, Yitong and Ma, Xingjun and Jin, Jiong and Yu, Han and Ng, Kee Siong},
  journal={IEEE Transactions on Parallel and Distributed Systems},
  volume={31},
  number={11},
  pages={2524--2541},
  year={2020},
  publisher={IEEE}
}

@inproceedings{bhagoji2019analyzing,
  title={Analyzing federated learning through an adversarial lens},
  author={Bhagoji, Arjun Nitin and Chakraborty, Supriyo and Mittal, Prateek and Calo, Seraphin},
  booktitle={International conference on machine learning},
  pages={634--643},
  year={2019},
  organization={PMLR}
}

@inproceedings {247652,
author = {Minghong Fang and Xiaoyu Cao and Jinyuan Jia and Neil Gong},
title = {Local Model Poisoning Attacks to {Byzantine-Robust} Federated Learning},
booktitle = {29th USENIX Security Symposium (USENIX Security 20)},
year = {2020},
isbn = {978-1-939133-17-5},
pages = {1605--1622},
url = {https://www.usenix.org/conference/usenixsecurity20/presentation/fang},
publisher = {USENIX Association},
month = aug
}

@inproceedings{zhao2025sok,
  title={Sok: Watermarking for ai-generated content},
  author={Zhao, Xuandong and Gunn, Sam and Christ, Miranda and Fairoze, Jaiden and Fabrega, Andres and Carlini, Nicholas and Garg, Sanjam and Hong, Sanghyun and Nasr, Milad and Tramer, Florian and others},
  booktitle={2025 IEEE Symposium on Security and Privacy (SP)},
  pages={2621--2639},
  year={2025},
  organization={IEEE}
}

@article{kingma2014adam,
  title={Adam: A method for stochastic optimization},
  author={Kingma, Diederik P and Ba, Jimmy},
  journal={arXiv preprint arXiv:1412.6980},
  year={2014}
}

@misc{eval-harness,
  author       = {Gao, Leo and Tow, Jonathan and Abbasi, Baber and Biderman, Stella and Black, Sid and DiPofi, Anthony and Foster, Charles and Golding, Laurence and Hsu, Jeffrey and Le Noac'h, Alain and Li, Haonan and McDonell, Kyle and Muennighoff, Niklas and Ociepa, Chris and Phang, Jason and Reynolds, Laria and Schoelkopf, Hailey and Skowron, Aviya and Sutawika, Lintang and Tang, Eric and Thite, Anish and Wang, Ben and Wang, Kevin and Zou, Andy},
  title        = {The Language Model Evaluation Harness},
  month        = 07,
  year         = 2024,
  publisher    = {Zenodo},
  version      = {v0.4.3},
  doi          = {10.5281/zenodo.12608602},
  url          = {https://zenodo.org/records/12608602},
}

@inproceedings{bottou2010large,
  title={Large-Scale Machine Learning with Stochastic Gradient Descent},
  author={Bottou, L{\'e}on},
  booktitle={Proceedings of COMPSTAT'2010},
  pages={177--187},
  year={2010},
  publisher={Springer},
}

@inproceedings{yao2026federated,
  title={Federated large language models: Current progress and future directions},
  author={Yao, Yuhang and Zhang, Jianyi and Wu, Junda and Huang, Chengkai and Xia, Yu and Yu, Tong and Zhang, Ruiyi and Kim, Sungchul and Rossi, Ryan and Li, Ang and others},
  booktitle={Pacific-Asia Conference on Knowledge Discovery and Data Mining},
  pages={203--221},
  year={2026},
  organization={Springer},
}

@inproceedings{allouah2023fixing,
  title={Fixing by mixing: A recipe for optimal byzantine ml under heterogeneity},
  author={Allouah, Youssef and Farhadkhani, Sadegh and Guerraoui, Rachid and Gupta, Nirupam and Pinot, Rafa{\"e}l and Stephan, John},
  booktitle={International Conference on Artificial Intelligence and Statistics},
  pages={1232--1300},
  year={2023},
  organization={PMLR}
}

@article{hsu2019measuring,
  title={Measuring the effects of non-identical data distribution for federated visual classification},
  author={Hsu, Tzu-Ming Harry and Qi, Hang and Brown, Matthew},
  journal={arXiv preprint arXiv:1909.06335},
  year={2019}
}

@article{pearlmutter1994fast,
    author = {Pearlmutter, Barak A.},
    title = {Fast Exact Multiplication by the Hessian},
    journal = {Neural Computation},
    volume = {6},
    number = {1},
    pages = {147-160},
    year = {1994},
    month = {01},
    issn = {0899-7667},
    doi = {10.1162/neco.1994.6.1.147},
    url = {https://doi.org/10.1162/neco.1994.6.1.147},
    eprint = {https://direct.mit.edu/neco/article-pdf/6/1/147/812672/neco.1994.6.1.147.pdf},
}

@article{baruch2019little,
  title={A little is enough: Circumventing defenses for distributed learning},
  author={Baruch, Gilad and Baruch, Moran and Goldberg, Yoav},
  journal={Advances in Neural Information Processing Systems},
  volume={32},
  year={2019}
}

@inproceedings{shejwalkar2021manipulating,
  title={Manipulating the byzantine: Optimizing model poisoning attacks and defenses for federated learning},
  author={Shejwalkar, Virat and Houmansadr, Amir},
  booktitle={NDSS},
  year={2021}
}

@article{geiping2020witches,
  title={Witches' brew: Industrial scale data poisoning via gradient matching},
  author={Geiping, Jonas and Fowl, Liam and Huang, W Ronny and Czaja, Wojciech and Taylor, Gavin and Moeller, Michael and Goldstein, Tom},
  journal={arXiv preprint arXiv:2009.02276},
  year={2020}
}
 
\appendices
\section{Open Science}
This paper has an accompanying artifact package intended to support review, reproducibility, and future comparison. The artifact contains the code, configuration files, scripts, and documentation needed to reproduce the core experimental results reported in the paper, including the evaluation of existing LLM watermarks under federated learning, the \fedwm optimization procedure, robust aggregation experiments, post-training removal attacks, and model-utility measurements. The anonymous repo is available at \url{https://anonymous.4open.science/r/provenance-D89F/}.

\myparatight{Artifacts provided}
We will provide the following artifacts for double-blind review:

\begin{itemize}
    \item \myparatight{Source code} The implementation of the federated fine-tuning pipeline, including FedAvg, RandEigen-based robust aggregation, client-side training, server-side optimization, LoRA fine-tuning, and the \fedwm watermark-camouflage procedure.
    \item \myparatight{Watermarking and detection code} Scripts for generating watermarked data using the evaluated LLM watermarking schemes, including KGW+, AAR, KTH-ITS, and KTH-EXP-style detection, together with scripts for computing post-training detection $p$-values.
    \item \myparatight{Experiment configurations} Configuration files specifying the model families, datasets, number of clients, watermarking-client ratios, communication rounds, optimizer settings, LoRA parameters, batch sizes, watermark parameters, and random seeds used in the experiments.
    \item \myparatight{Evaluation scripts} Scripts for reproducing the main tables and figures, including watermark detectability under FedAvg and RandEigen, \fedwm detectability, ablation studies, post-training removal attacks, evasion-rate measurements, and model-utility evaluations.
\end{itemize}

\myparatight{Datasets and models}
The experiments use publicly available models and datasets, including the model families and datasets described in the evaluation section, such as Pythia, Qwen, LLaMA, Alpaca, OpenWebText, OASST1, C4, and MMLU-style evaluation data. Because some models and datasets are distributed under third-party licenses, we will not redistribute their raw contents in the artifact package. Instead, the artifact will include download instructions, preprocessing scripts, dataset-format specifications, and configuration files identifying the exact resources used. Reviewers can obtain the underlying datasets and pretrained models from their official sources subject to their respective licenses.

\section{Experimental Details}
\subsection{Seed Selection} \label{append:seed}
\begin{table}[t]
\centering
\caption{Seeds used by each watermarking client in the baseline experiments (Table~\ref{tab:baseline_prev} and Table~\ref{tab:baseline_opt}). Seeds are selected per dataset, model, and watermark scheme. The \emph{same} seed assignment is reused across all aggregation methods.}
\label{tab:seeds-table2-3}
\begin{tabular}{llccc}
\toprule
Dataset & WM & \multicolumn{3}{c}{Table II} \\
        &    & Pythia & Qwen & LLaMA \\
\midrule
\multirow{4}{*}{Alpaca}
 & KGW+     & (3, 6) & (3, 6) & (3, 6) \\
 & AAR      & (0, 1) & (0, 1) & (0, 1) \\
 & KTH-EXP  & (0, 1) & (0, 1) & (0, 1) \\
 & KTH-ITS  & (0, 1) & (0, 1) & (0, 1) \\
\midrule
\multirow{4}{*}{OpenWebText}
 & KGW+     & (3, 6) & (3, 6) & (3, 6) \\
 & AAR      & (4, 5) & (4, 5) & (4, 5) \\
 & KTH-EXP  & (4, 5) & (4, 5) & (4, 5) \\
 & KTH-ITS  & (4, 5) & (4, 5) & (4, 5) \\
\bottomrule
\end{tabular}

\vspace{1em}

\begin{tabular}{llccc}
\toprule
Dataset & WM & \multicolumn{3}{c}{Table III} \\
        &    & Pythia & Qwen & LLaMA \\
\midrule
\multirow{1}{*}{Alpaca}
 & CamoMark & (3, 6) & (3, 6) & (3, 6) \\
\midrule
\multirow{1}{*}{OpenWebText}
 & CamoMark & (3, 6) & (3, 6) & (8, 6) \\
\bottomrule
\end{tabular}
\end{table}

\begin{table*}[t]
\centering
\caption{Seeds used by each watermarking client for the watermark-ratio ablation experiments on \alpaca\ under \randeigen\ (Table~\ref{tab:diffR}). For each model, we report the seeds used by \Kirchenbauer\ and by \fedwm\ side by side.}
\label{tab:seeds-table4}
\begin{tabular}{ll cc cc cc}
\toprule
 & & \multicolumn{2}{c}{Pythia} & \multicolumn{2}{c}{Qwen} & \multicolumn{2}{c}{LLaMA} \\
\cmidrule(lr){3-4} \cmidrule(lr){5-6} \cmidrule(lr){7-8}
WM Ratio & Client \# & KGW+ & CamoMark & KGW+ & CamoMark & KGW+ & CamoMark \\
\midrule
\multirow{2}{*}{6.6\% ($k$=2)}
 & 1 & 3 & 3 & 3 & 3 & 3 & 3 \\
 & 2 & 6 & 6 & 6 & 6 & 6 & 6 \\
\midrule
\multirow{3}{*}{10.0\% ($k$=3)}
 & 1 & 3 & 3 & 3 & 3 & 3 & 3 \\
 & 2 & 6 & 6 & 6 & 6 & 6 & 6 \\
 & 3 & 9 & 9 & 9 & 9 & 9 & 9 \\
\midrule
\multirow{4}{*}{13.3\% ($k$=4)}
 & 1 & 3 & 3 & 3 & 3 & 3 & 3 \\
 & 2 & 6 & 6 & 6 & 6 & 6 & 6 \\
 & 3 & 7 & 9 & 7 & 7 & 7 & 7 \\
 & 4 & 8 & 7 & 8 & 8 & 8 & 8 \\
\bottomrule
\end{tabular}
\end{table*}

In this section, we report the seeds assigned to each watermarking client in Section~\ref{sec:eval}. Table~\ref{tab:seeds-table2-3} lists the seeds used Table~\ref{tab:baseline_prev} and Table~\ref{tab:baseline_opt}, while Table~\ref{tab:seeds-table4} lists the seeds used for Table~\ref{tab:diffR}. In all cases, seeds are selected such that the pre-fine-tuning detection $p$-value is approximately 0.5, i.e., matching the null hypothesis. It intends to equalize each experiment's starting detection statistic before any federated training occurs, so that differences in post-training $\pval$ across models, datasets, watermarking schemes, aggregators, and watermark ratios can be attributed to the experimental factor being varied rather than to an accidentally favorable or unfavorable starting condition.

The pre-training $p$-value obtained for a given seed is stochastic, based on the LLM watermarking algorithm and the generated text itself, and it is orthogonal to our approach \fedwm. Our empirical observation suggests that null hypothesis (pre-training) $p$-values of prior LLM watermarking works should be carefully reported.

\subsection{Clean Client Distribution: Natural vs. Synthetic} \label{append:cmp}
\begin{table}[t]
\centering
\small
\caption{Filtering effectiveness of $\randeigen$ under different clean-client data distributions for $\pythia$ trained on $\alpaca$. $\Kirchenbauer$ is heavily filtered regardless of the clean data distribution, whereas $\EXP$ is heavily filtered when the clean clients use natural data.}
\label{tab:baseline_cmp}
\begin{tabular}{ccc}
\toprule
\textbf{WM} & \textbf{Clean Client}  & \textbf{$\bm{\evasionrate}$} \\
& \textbf{Data} & (\%) \\
\midrule
$\Kirchenbauer$     & synthetic            & 7.29 \\
                    & natural              & 2.08 \\
\midrule
$\EXP$              & synthetic            & 59.38  \\
                    & natural              & 3.13 \\
\bottomrule
\end{tabular}
\end{table}

Table~\ref{tab:baseline_cmp} reports the evasion rate ($\evasionrate$) of $\randeigen$ across different clean-client data distributions and watermarking schemes for $\pythia$ trained on $\alpaca$. A lower $\evasionrate$ indicates that more watermarked updates were filtered by the aggregator, leading to lower detection confidence after FL. The table shows that $\Kirchenbauer$ achieves a low $\evasionrate$ of 7.29\% and 2.08\% regardless of the clean-client data distribution. This is expected, since $\Kirchenbauer$ introduces a distortion into the output distribution, allowing the aggregator to easily distinguish watermarked updates from clean updates. In contrast, $\EXP$ achieves a much higher $\evasionrate$ of 59.38\% when the clean clients use a synthetic dataset generated from the same underlying model and data source. This is also expected, because $\EXP$ is practically distortion-free: it introduces no distribution shift into the watermarked dataset, and therefore no distinguishable difference between watermarked and clean synthetic updates. However, when the clean clients instead use the original natural dataset, in which the clean data has not been passed through the generation model, $\evasionrate$ drops back down to 3.13\%. This occurs because of the inherent distributional gap between synthetic and natural data, which exposes a key limitation of prior work: a desirable watermark should evade the aggregator's filtering regardless of the clean client data distribution, but all existing methods evaluated fail to satisfy this property. This motivates the need for a more reliable and robust watermark, $\fedwm$.

\subsection{Ablation Study for $\coeReg$} \label{append:ablation-coeReg}
\begin{table}[t]
\centering
\caption{Ablation study on $\coeReg$ for $\pythia$ trained on $\alpaca$. $\fedwm$ exhibits limited sensitivity to $\coeReg$: both $\evasionrate$ and $\pval$ vary only slightly as $\coeReg$ increases from 0 to 10.}
\label{tab:ablation-coeReg}
\begin{tabular}{lcc}
\toprule
\textbf{$\bm{\coeReg}$} & \textbf{Evasion Rate} & \textbf{$\pval$} \\
\midrule
0    & 20.8\% & $6.75 \times 10^{-2}$ \\
0.1  & 19.8\% & $4.28 \times 10^{-2}$ \\
1    & 19.8\% & $1.69 \times 10^{-2}$ \\
10   & 19.8\% & $3.43 \times 10^{-2}$ \\
\bottomrule
\end{tabular}
\end{table}

$\coeReg$ controls the trade‑off between gradient camouflage ($\lossRec$) and watermark preservation ($\lossReg$). We conduct an ablation study varying $\coeReg \in \{0, 0.1, 1, 10\}$ on $\pythia$ trained on $\alpaca$. Table~\ref{tab:ablation-coeReg} shows that $\fedwm$ exhibits limited sensitivity to $\coeReg$: the first-round evasion rate varies by at most 1\%, and the final p-value ranges from $1.69 \times 10^{-2}$ to $6.75 \times 10^{-2}$. 

Varying $\coeReg$ further allows us to show that cosine similarity at the input embedding level is closely tied to the model's output-token statistics, which in turn affect watermark detection confidence. This is a core principle underlying $\fedwm$. To demonstrate, we conduct a small-scale ablation on $\pythia$ trained on a single batch of $\alpaca$. We vary $\coeReg$ and use $\fedavg$ as the aggregation method to isolate the effect from any filtering introduced by robust aggregation. As $\coeReg$ increases from 0 to 10, the cosine similarity between the optimized and initial watermark embeddings rises from 0.9909 to 0.9979, and detection confidence correspondingly improves, with $\pval$ decreasing from $6.78 \times 10^{-3}$ to $7.73 \times 10^{-5}$. This positive correlation provides direct evidence that preserving embedding similarity also preserves output-level watermark detectability.

\section{Complexity and Theoretical Analysis of \fedwm}
\label{append:theory}
\subsection{Computational Complexity}
\label{append:complexity}
For each watermarked batch, $\fedwm$ first computes the clean proxy gradient at cost $\costG$, the cost of one forward--backward pass through the LoRA-adapted model, and then runs $K$ optimization iterations (Algorithm~\ref{alg:camomark}). Each iteration computes the model gradient at cost $\costG$ and the embedding gradient at cost $\costH$, the cost of computing a mixed Hessian--vector product. After the loop terminates and $\embedsBest$ has been selected, we recompute the final gradient once more at cost $\costG$, giving a per-batch cost of $2\costG + K(\costG + \costH)$. Since a Hessian--vector product costs asymptotically the same as a gradient computation, $\costH = \mathcal{O}(\costG)$~\cite{pearlmutter1994fast}, and the total complexity over $B$ batches is therefore $\mathcal{O}(B \cdot K \cdot \costG)$. Empirically, on $\pythia$ one optimization iteration takes 0.40 seconds, so at $K=50$, optimization takes at most 20 seconds per batch. We also optimize only once every 5 or 10 rounds rather than every round, further reducing overhead.

\subsection{Theoretical Discussion: Evasion Optimality}
\label{append:evasion-theory}
$\randeigen$ preferentially filters more deviated updates via a covariance-based convergence criterion, achieving the information-theoretically optimal bias bound $\mathcal{O}(\sqrt\epsilon)$~\cite{choudhary2024attacking,lee2025practical}. Given the true clean mean and covariance, a client could construct maximally deviated inlier updates, bypassing filtering. Lacking these statistics, $\fedwm$ estimates them through a proxy model, achieving best-case inlier behavior when the reconstruction loss is minimized and the estimates are accurate. 

While optimal evasion admits this theoretical characterization, provably learning a detectable watermark is a separate problem. To our knowledge, there is currently no theoretical characterization of how much residual watermark signal guarantees successful detection after training. Prior work evaluates post-training watermark detectability only empirically~\cite{sander2024watermarking,gu2023learnability}, and we follow the same practice: our experiments (Section~\ref{sec:eval}) show that $\fedwm$'s regularization preserves sufficient watermark signal for low \pval.

\section{Details on LLM Watermarking}
\label{append:llm-watermarking}
In this section, we present the detailed generation and detection procedures for the three representative watermarking families evaluated in this paper: \Kirchenbauer, \Aarson, and \Kuditipudi.

\subsection{\Kirchenbauer~\cite{kirchenbauer2023watermark}}
\label{app:kgw}

$\Kirchenbauer$ is a hashing-based green-list watermark. During watermark generation, when generating the $i$th token, the scheme hashes the previous $\ngrams$ tokens together with a secret key $\secretkey$ to seed a pseudorandom number generator. Using this generator, the vocabulary $\vocab$ is partitioned into a green list $G_i$ of size $\greenratio|\vocab|$ and its complement, the red list $R_i$. The logits of tokens in $G_i$ are then increased by $\logitsbias>0$ before the softmax, producing a biased next-token distribution from which the $i$th token is sampled.

For watermark detection, let $X_N=(x_1,\dots,x_N)$ denote the effective token sequence used for detection after any preprocessing, so that $N$ is the number of tokens scored by the detector. For each token $x_i$, define the per-token indicator
\begin{equation}
  \operatorname{WScore}\bigl(
    x_i;\secretkey,(x_{i-1},\dots,x_{i-\ngrams})
  \bigr)
  =
  \mathbb{1}\{x_i\in G_i\}.
\end{equation}
The cumulative score is
\begin{equation}
  S(X_N)
  =
  \sum_{i=1}^{N}
  \operatorname{WScore}\bigl(
    x_i;\secretkey,(x_{i-1},\dots,x_{i-\ngrams})
  \bigr),
\end{equation}
which counts the number of green tokens in the effective text.

Under $\nullhypo$, the candidate text is generated independently of the secret key. Under the standard idealization that the keyed hash behaves as a pseudorandom function, the indicators are i.i.d.\
$\operatorname{Bernoulli}(\greenratio)$. Consequently,
\begin{equation}
  S(X_N)
  \sim
  \operatorname{Binomial}(N,\greenratio)
  \qquad\text{under }\nullhypo.
\end{equation}
Watermark detection computes the exact one-sided upper-tail \pval
\begin{equation}
  \pval
  =
  \Pr\!\left[
    S'\geq S(X_N)
    \,\middle|\,
    S'\sim\operatorname{Binomial}(N,\greenratio)
  \right].
\end{equation}
A smaller \pval gives stronger evidence that the text is watermarked. The parameters $\greenratio\in(0,1)$ and $\logitsbias>0$ control the trade-off between watermark detectability and generation quality.

\subsection{\Aarson~\cite{aaronson2023watermarking}}
\label{app:aar}

$\Aarson$ shares the same context-hashing procedure as $\Kirchenbauer$ (Appendix~\ref{app:kgw}). At step $i$, the previous $\ngrams$ tokens and the secret key $\secretkey$ seed a pseudorandom number generator. Rather than partitioning the vocabulary into green and red lists, $\Aarson$ uses the generator to sample via the Gumbel-max trick. For every vocabulary item $v\in\vocab$, the generator produces a value $r_i^{(v)}\in(0,1)$, with the values $\{r_i^{(v)}\}_{v\in\vocab}$ modeled as i.i.d.\
$\operatorname{Uniform}(0,1)$ random variables. The next token is selected as
\begin{equation}
  x_i
  =
  \arg\max_{v\in\vocab}
  \left(r_i^{(v)}\right)^{1/p_i^{(v)}},
\end{equation}
where $p_i^{(v)}$ is the model's non-watermarked probability of token $v$ at step $i$.

When a fresh independent pseudorandom vector is used at every position, this rule samples exactly from $p_i$:
\begin{equation}
  \Pr[x_i=v]=p_i^{(v)}
  \qquad
  \text{for every }v\in\vocab.
\end{equation}
Under this condition, the resulting sequence has the same distribution as text produced by non-watermarked, original model sampling. However, when the hash context repeats, it introduces dependencies between positions and can distort the joint distribution over generated text. Since natural language is highly repetitive, such context repetition occurs often in practice, which is why $\Aarson$ is only theoretically, and not always practically, distortion-free.

For watermark detection, again let $X_N=(x_1,\dots,x_N)$ denote the effective token sequence used for detection after any preprocessing. The detector recomputes $r_i^{(x_i)}$ at each position using the same secret key and hash function as generation, and forms the statistic
\begin{equation}
  T(X_N)
  =
  -\sum_{i=1}^{N}
  \log\left(1-r_i^{(x_i)}\right).
\end{equation}
Under $\nullhypo$, the values $r_i^{(x_i)}$ are modeled as i.i.d.\
$\operatorname{Uniform}(0,1)$ random variables. Therefore,
$-\log(1-r_i^{(x_i)})\sim\operatorname{Exponential}(1)$ and
\begin{equation}
  T(X_N)\sim\operatorname{Gamma}(N,1)
  \qquad\text{under }\nullhypo.
\end{equation}
Watermark detection computes the one-sided upper-tail \pval
\begin{equation}
  \pval
  =
  \Pr\!\left[
    T'\geq T(X_N)
    \,\middle|\,
    T'\sim\operatorname{Gamma}(N,1)
  \right].
\end{equation}
A smaller \pval gives stronger evidence that the text is watermarked.

\subsection{\Kuditipudi~\cite{kuditipudi2023robust}}
\label{app:kth}

$\Kuditipudi$ is a distortion-free key-sequence watermark. Rather than deriving randomness by hashing the local context, it predefines a random key sequence $\xi=(\xi_1,\dots,\xi_n)$ that governs how the model distribution $p^{(i)}$ is mapped to the next token. Marginalizing over the random key, and assuming that key elements are not reused, the generated text follows the original non-watermarked model distribution.

The scheme proposes two decoders: inverse-transform sampling ($\ITS$) and exponential-minimum sampling ($\EXP$). $\EXP$ uses a sampling rule equivalent to the Gumbel-max construction underlying $\Aarson$ (Appendix~\ref{app:aar}), but obtains its randomness from the pre-generated key sequence rather than by hashing the local context.

For watermark detection, let $X_N=(x_1,\dots,x_N)$ denote the effective token sequence used for detection after any preprocessing. Given the secret key sequence $\xi$, the detector searches over possible alignments and computes
\begin{equation}
  \phi(X_N,\xi)
  =
  \min_j
  d\left(
    X_N,
    \xi_{j:j+N-1}^{\mathrm{cyclic}}
  \right),
\end{equation}
where $d$ is a decoder-specific alignment cost and
$\xi_{j:j+N-1}^{\mathrm{cyclic}}$ denotes a length-$N$ segment of the key sequence beginning at position $j$, with cyclic indexing. A smaller value of $\phi$ indicates a better alignment between the candidate text and the key sequence.

Under $\nullhypo$, the candidate text and the secret key are independent. The \pval is estimated by comparing the observed alignment cost with those obtained from $B$ independently sampled null keys:
\begin{equation}
  \widehat{\pval}
  =
  \frac{
    1+
    \sum_{b=1}^{B}
    \mathbb{1}\left\{
      \phi(X_N,\xi^{(b)})
      \leq
      \phi(X_N,\xi)
    \right\}
  }{B+1}.
\end{equation}
Smaller alignment costs, and hence smaller \pvals, provide stronger evidence that the text is watermarked.

\end{document}